\documentclass[12pt]{article}
\usepackage[utf8]{inputenc}
\usepackage[margin = 1in]{geometry}
\usepackage{amsmath,amsthm,amssymb}
\usepackage{setspace}
\usepackage{enumerate}
\usepackage{natbib}
\usepackage[dvipsnames]{xcolor}
\usepackage{mathrsfs}
\usepackage{blindtext}
\usepackage{graphicx}
\usepackage{bm}
\usepackage{mathrsfs}
\usepackage{enumitem}
\usepackage{algorithm}
\usepackage{algorithmicx}
\usepackage{algpseudocode}

\newcommand{\mythanks}[1]{\thanks{\parbox[t]{0.95\textwidth}{#1}}}

\newif\ifarxiv
\arxivtrue

\definecolor{alertcolor}{RGB}{92, 16, 127}
\definecolor{tonycolor}{RGB}{232, 55, 70}

\title{
    The How and Why of Bayesian Nonparametric Causal Inference
}
\author{
  Antonio R. Linero\mythanks{Department of Statistics and Data Sciences, University of Texas at Austin, email: \url{antonio.linero@austin.utexas.edu}}
\and 
  Joseph L. Antonelli\mythanks{Department of Statistics, University of Florida, email: \url{jantonelli@ufl.edu}}}

\newcommand{\ba}{\bm a}
\newcommand{\bA}{\bm A}

\newcommand{\bx}{\bm x}
\newcommand{\bX}{\bm X}
\newcommand{\bY}{\bm Y}

\newcommand{\Bernoulli}{\operatorname{Bernoulli}}
\newcommand{\Beta}{\operatorname{Beta}}
\newcommand{\Categorical}{\operatorname{Categorical}}
\newcommand{\Cauchy}{\operatorname{Cauchy}}
\newcommand{\Cor}{\operatorname{Cor}}

\newcommand{\Dirichlet}{\operatorname{Dirichlet}}
\newcommand{\E}{\mathbb E}

\newcommand{\GP}{\operatorname{GP}}
\newcommand{\Identity}{\mathrm I}
\newcommand{\iid}{\stackrel{\textnormal{iid}}{\sim}}
\newcommand{\indep}{\stackrel{\textnormal{indep}}{\sim}}
\newcommand{\med}{\operatorname{median}}
\newcommand{\Multinomial}{\operatorname{Multinomial}}
\newcommand{\Normal}{\operatorname{Normal}}
\newcommand{\Poisson}{\operatorname{Poisson}}
\newcommand{\Reals}{\mathbb R}
\newcommand{\sC}{\mathscr C}

\newcommand{\sM}{\mathcal M}
\newcommand{\sX}{\mathscr X}

\newcommand{\Tree}{\mathcal T}
\newcommand{\Treef}{\operatorname{Tree}}
\newcommand{\Polya}{Pólya}
\newcommand{\Var}{\operatorname{Var}}

\theoremstyle{definition}
\newtheorem{definition}{Definition}

\usepackage[colorlinks, citecolor = NavyBlue]{hyperref}
\date{}

\begin{document}

\maketitle

\begin{abstract}
    Spurred on by recent successes in causal inference competitions, Bayesian nonparametric (and high-dimensional) methods have recently seen increased attention in the causal inference literature. In this paper, we present a comprehensive overview of Bayesian nonparametric applications to causal inference. Our aims are to (i) introduce the fundamental Bayesian nonparametric toolkit; (ii) discuss how to determine which tool is most appropriate for a given problem; and (iii) show how to avoid common pitfalls in applying Bayesian nonparametric methods in high-dimensional settings. Unlike standard fixed-dimensional parametric problems, where outcome modeling alone can sometimes be effective, we argue that most of the time it is necessary to model both the selection and outcome processes.
\end{abstract}

\doublespacing

\section{Introduction}

Bayesian nonparametric methods have recently seen increased interest in the causal inference literature. Much of this interest is due to the success of methods based on Bayesian additive regression trees in the annual American Causal Inference Conference (ACIC) competitions \citep{dorie2019automated, hill2011bayesian, hahn2020bayesian}, although there have also been successful applications of Bayesian nonparametric methods based on Gaussian processes \citep{ray2019debiased, ren2021bayesian, vegetabile2020optimally, branson2019nonparametric} and infinite latent-class models \citep{kim2016framework, roy2018bayesian, kim2019bayesian}. In general, Bayesian nonparametrics offers both the flexibility of modern machine learning algorithms and the statistically-principled uncertainty quantification of Bayesian inference. That this is possible is evinced by the strong performance of appropriately-designed Bayesian nonparametric methods in terms of estimation accuracy and coverage of credible intervals \citep{dorie2019automated, roy2018bayesian, xu2018bayesian}, as well as some promising theoretical results \citep{ray2020semiparametric}.

Applying Bayesian nonparametrics to causal inference is rarely as simple as taking an off-the-shelf nonparametric prior and applying it in the same way one would to a prediction problem. Causal inference problems are often targeted in the sense that the final aim is to estimate a low-dimensional parameter $\psi(\theta)$, with nonparametric techniques used to deal with high or infinite dimensional nuisance parameters. As discussed in Section~\ref{sec:prior-dogmatism}, the shrinkage induced by nonparametric models on the causal estimands introduces subtle, but serious, complications. Because of this, special care should be taken when applying Bayesian nonparametrics.

The goal of this paper is to provide a guide to Bayesian nonparametric causal inference for both (a) causal inference researchers interested in applying Bayesian nonparametrics and (b) those familiar with Bayesian nonparametrics who are interested in tackling causal inference problems. A textbook-level treatment of this topic is given by the forthcoming monograph of \citet{daniels2021bayesian}. In Section~\ref{sec:BNPpriors} we provide an introduction to the fundamental tools of Bayesian nonparametrics such as Gaussian processes, Dirichlet processes, Bayesian additive regression trees, and spike-and-slab priors, which will allow readers to design Bayesian nonparametric methods to suit their needs. In Section~\ref{sec:estimands} we provide an overview of the potential outcomes framework under strong ignorability, discuss possible estimands of interest, and discuss the assumptions required to identify those estimands. In Section~\ref{sec:bayesian-causal-inference} we discuss the causal-inference-specific modifications needed to apply Bayesian nonparametrics methods, show how to compute causal parameters from a posterior sample using Monte Carlo integration, and discuss extensions to settings with high-dimensional confounders. In Section~\ref{sec:OtherEstimands} we discuss extensions of Bayesian nonparametrics to different estimands, such as direct and indirect effects in causal mediation analysis and local treatment effects in regression discontinuity designs. We highlight these approaches in an application to estimate the effect of smoking on medical expenditures in Section~\ref{sec:applications}. We 
close in Section~\ref{sec:discussion} with a discussion.

\section{Priors on Functions and Distributions}
\label{sec:BNPpriors}

Nonparametric methods typically estimate function-valued parameters such as regression or density functions. In order to apply Bayesian inference to nonparametric modeling we therefore must consider priors $\Pi(d\theta)$ where the parameter $\theta$ is itself a function. While there are many such priors, the following are the most important in our experience: Gaussian processes, Bayesian additive regression trees, and Dirichlet processes. Textbook level treatments of both the theory and practice of Bayesian nonparametric methods are given by \citet{ghosal2017fundamentals} and \citet{muller2015bayesian}, respectively.

Throughout this section, we consider a non-causal setup where we have a covariate vector $X_i \in \Reals^P$ and outcome $Y_i \in \Reals$, and we are interested in learning the conditional distribution $f_\theta(y \mid x)$ of $Y_i$ given $X_i$. We will let $\bX$ denote the matrix with $i^{\text{th}}$ row given by $X_i^\top$ and let $\bY = (Y_1, \ldots, Y_N)$.

\subsection{Aside on Fitting Bayesian Models}

A common strategy for fitting Bayesian models is to sample approximately from the posterior distribution $\Pi(d\theta \mid \bX, \bY)$ using \emph{Markov chain Monte Carlo} (MCMC). In an ideal setting, we can obtain a sample from the posterior distribution $\theta_1, \ldots, \theta_B \iid \Pi(d\theta \mid \bX, \bY)$ for some large $B$ and then use these samples to approximate posterior means and credible intervals; for example, to generate a credible interval for a parameter $\psi(\theta)$ we might use $[\psi_{0.25}, \psi_{0.975}]$ where $\psi_\alpha$ is the $\alpha^{\text{th}}$ quantile of the $\psi(\theta_b)$'s. In practice, sampling exactly from $\Pi(d\theta\mid\bX, \bY)$ will usually be impossible; however, it is usually easy to construct a sequence of \emph{dependent} $\theta_b$'s which converge in distribution to $\Pi(d\theta\mid\bX, \bY)$ using MCMC. We do not attempt to cover the nuances of MCMC here, and refer readers to \citet{robert2010introducing} for a detailed treatment of this topic. The only essential fact we make use of here is that we can obtain samples $\theta_1, \ldots, \theta_B$ using \emph{some} MCMC scheme.

\subsection{Gaussian Processes}
\label{sec:GPs}

Consider the semiparametric regression model which takes 
\begin{align}
    \label{eq:semiparametric}
    Y_i = g(X_i) + \epsilon_i, \qquad \epsilon_i \iid \Normal(0, \sigma^2)
\end{align}
with $g(x)$ being an arbitrary function $g: \Reals^P \to \Reals$. A desirable property for any prior $g \sim \Pi$ is that the support of $\Pi$ is ``large'' in the sense that, for any continuous $g_0$, the prior assigns positive mass to any neighborhood of $g_0$. For example, we might ask that $\Pi(\sup_{x \in C} |g(x) - g_0(x)| \le \epsilon) > 0$ for all $\epsilon$ and for all compact sets $C \subseteq \mathbb R^P$; this ensures that $g$ and $g_0$ are uniformly close to each other on compact sets with positive probability. The simplest priors which accomplish this goal are \emph{Gaussian process} priors.


\begin{definition}
  Let $m(x)$ and $K(x,x')$ be functions. We say that a random function $g: \Reals^P \to \Reals$ is a \emph{Gaussian process} with mean function $m(x)$ and covariance function $K(x,x')$ if, for every finite $D$ and collection of points $\bx = (x_1, \ldots, x_D)$ in $\Reals^P$, we have $(g(x_1), \ldots, g(x_D))^\top \sim \Normal\left(\mu, \Sigma\right)$ where $\mu = (m(x_1), \ldots, m(x_D))^\top$ and $\Sigma$ is a covariance matrix with $(j,k)^{\text{th}}$ element given by $K(x_j, x_k)$. To denote this we write $g \sim \GP(m, K)$
\end{definition}

For a detailed treatment of Gaussian processes see \citet{rasmussen2005gaussian}. Gaussian processes generalize the multivariate normal distribution from vectors to functions, with $m$ and $K$ playing the role of the mean and covariance matrix respectively. In order for this prior to be valid it is necessary that $K(x,x')$ always produce a valid covariance matrix; functions with this property are called \emph{positive semi-definite}, or \emph{kernel}, functions.

\begin{figure}
    \centering
    \includegraphics[width=1\textwidth]{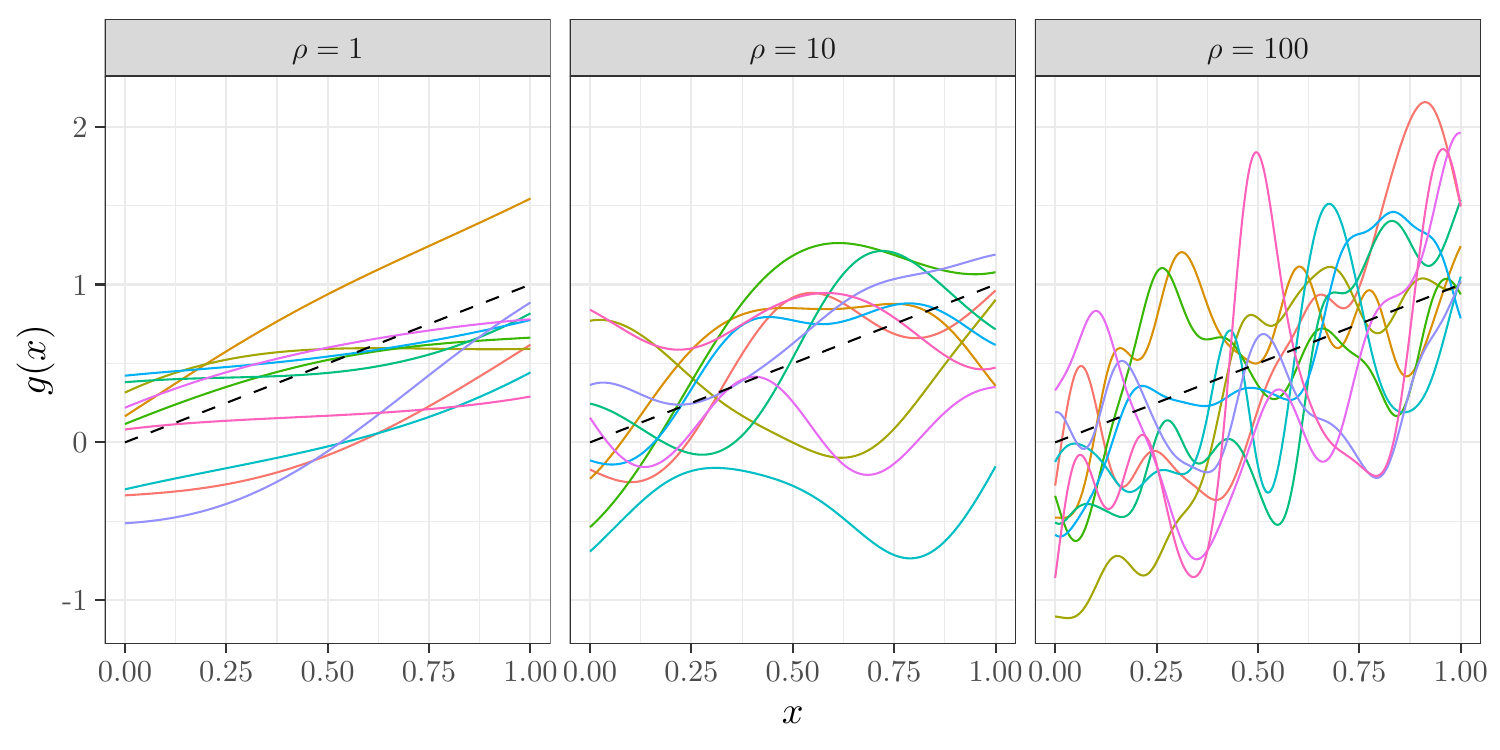}
    \caption{Samples of functions from a $\GP(m, K)$ distribution for $m(x) = x$ and $K(x,x') = 0.2 \, \exp\{-\rho(x-x')^2\}$ for $\rho \in \{1,10,100\}$. }
    \label{fig:gpplots}
\end{figure}

Samples of Gaussian processes in one dimension are given in Figure~\ref{fig:gpplots} with mean $m(x) = x$ and covariance given by the \emph{squared exponential kernel} $K(x,x') = \sigma_g^2 \, \exp\left\{- \rho (x-x')^2\right\}$ for $\rho \in \{1,10,100\}$ and $\sigma_g^2 = 0.2$. We see that the choice of the covariance function is extremely important, as it tells us how correlated the values of $g(x)$ and $g(x')$ are. Roughly speaking, the more correlated $g(x)$ is with $g(x + \Delta)$ for small $\Delta$'s, the smoother the $g(x)$ will be. For the squared exponential kernel, the parameter $\rho$ controls how far away $x$ and $x'$ must be for them to be nearly-uncorrelated, with larger values of $\rho$ corresponding to faster decreases in correlation, i.e., $\rho$ is analogous to a bandwidth parameter in kernel smoothing methods. When $\rho = 1$ we see that samples from the prior are very smooth on $[0,1]$, while when $\rho = 100$ the functions are extremely wiggly. In all cases, the samples from the prior are centered around $m(x)$; how close they are to $m(x)$ is controlled by $\sigma^2_g$, which is sometimes called the signal-to-noise parameter. 

Gaussian processes have conjugacy properties which make them convenient to work with. For the semiparametric regression model \eqref{eq:semiparametric} it turns out that the $\GP(m,K)$ distribution is a conjugate prior for $g$, with the posterior distribution given by $[g \mid \bX, \bY] \sim \GP(m^\star, K^\star)$ where 
\begin{align*}
    K^\star(\bx, \bx)
    &=
    K(\bx,\bx) - K(\bx, \bX) \left\{K(\bX,\bX) + \sigma^2 \, \Identity\right\}^{-1} K(\bX,\bx)
    \quad \text{and}
    \\
    m^\star(\bx) &= m(\bx) + K(\bx,\bX) \left\{K(\bX,\bX) + \sigma^2 \, \Identity\right\}^{-1} \{\bY - m(\bX)\},
\end{align*}
$m(\bx) = (m(x_1), \ldots, m(x_D))^\top$ and $K(\bx, \bx')$ is a matrix with $(j,k)^{\text{th}}$ entry $K(x_j, x'_k)$. Typically $m(x)$ and $K(x,x')$ will be indexed by some hyperparameters $h$ (e.g., the parameters $\sigma_g$ and $\rho$ for the squared exponential kernel) which should be learned from the data. These hyperparameters can be estimated using the marginal likelihood of the data
\begin{align}
    \label{eq:gp-marginal}
    f(\bY \mid \bX, \sigma^2, h)
    =
    \Normal\{\bY \mid m_h(\bX), K_h(\bX, \bX) + \sigma^2 \, \Identity\},
\end{align}
which can be derived by marginalizing out $g$. Fully-Bayesian inference on $(\sigma^2, h)$ can then be performed via Markov chain Monte Carlo (MCMC). Alternatively, we might estimate $(\sigma^2, h)$ by maximizing \eqref{eq:gp-marginal}; this strategy is known as \emph{empirical Bayes} in the statistics literature, or as \emph{type-II maximum likelihood} in the machine learning literature.

For models other than \eqref{eq:semiparametric}, Gaussian processes will not be conjugate and will require additional work to use. With slightly more effort they can be adapted to the nonparametric probit regression model $Y_i \sim \Bernoulli[\Phi\{g(X_i)\}]$ by using the data augmentation algorithm of \citet{albert1993bayesian} or to a logistic regression model using the \Polya-Gamma sampler of \citet{polson2013bayesian}. Alternatively, one can update $\bm g = (g(X_1), \ldots, g(X_N))^\top$ using Hamiltonian Monte Carlo \citep{neal2011mcmc} as implemented, for example, in the probabilistic programming language \texttt{Stan} \citep{rstan}.

The main difficulty with applying Gaussian processes in practice is that computing the posterior distribution ostensibly requires inverting the $N \times N$ matrix $K(\bX, \bX) + \sigma^2 \, \Identity$, a task which requires $O(N^3)$ computations. This is challenging for even moderately-sized problems (say, $N = 10,000$). Methods for reducing this computational burden are still being actively researched, though existing solutions include using low-rank approximations \citep{cressie2008fixed, banerjee2008gaussian} or choosing the kernel so that $K(\bX, \bX)$ is low rank, sparse, or otherwise conveniently structured for fast computations \citep{datta2016hierarchical,zhang2015full, katzfuss2021general}. Implementations include the \texttt{R} package \texttt{tgp} \citep{r-tgp} (which implements Gaussian process models, Bayesian decision tree models, and various combinations thereof) and the comprehensive \texttt{Matlab} package \texttt{gpstuff} \citep{vanhatalo2013gpstuff}.

\subsection{Bayesian Additive Regression Trees}

\begin{figure}
    \centering
    \includegraphics[width=.75\textwidth]{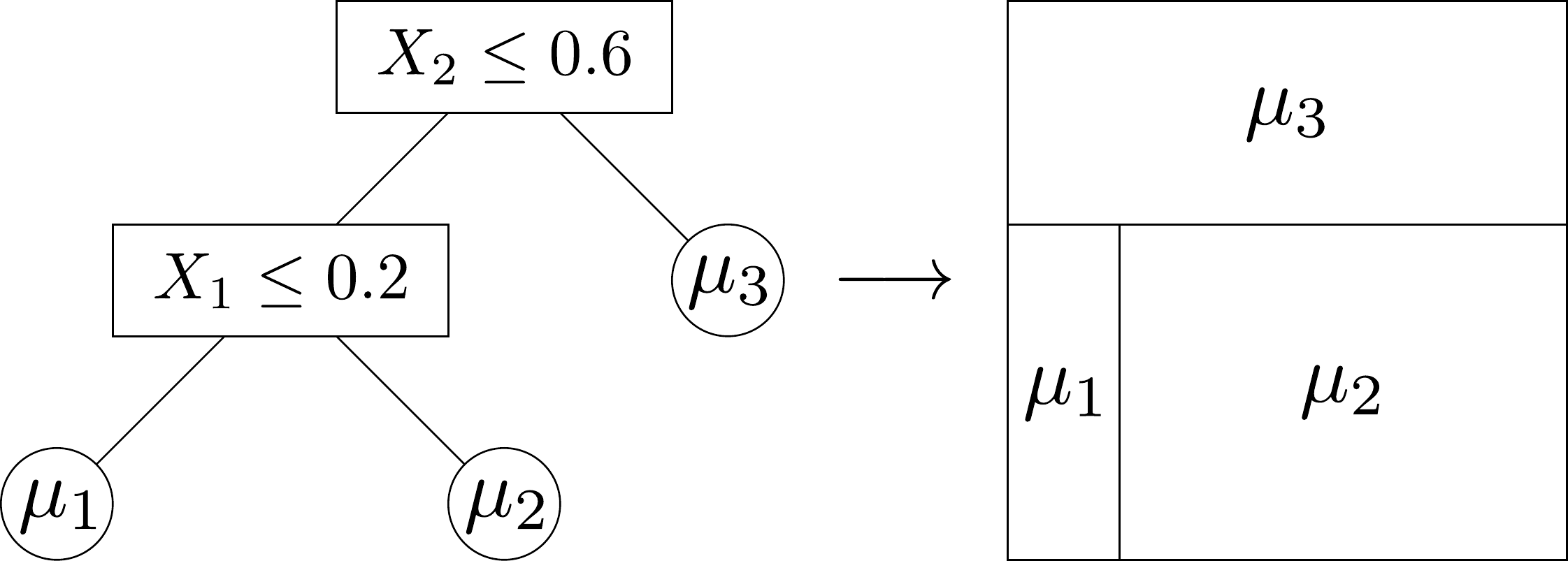}
    \caption{Schematic showing how a decision tree (left) produces a regression function $\Treef(x; \Tree, \sM)$ (right) which is a step function.}
    \label{fig:treefig}
\end{figure}

Like Gaussian processes, Bayesian additive regression tree (BART) models specify a prior on functions $g: \Reals^P \to \Reals$. BART takes $g(x)$ to be a \emph{sum of trees}
\begin{align}
    \label{eq:bart}
    g(x) = \sum_{t=1}^T \Treef(x; \Tree_t, \sM_t).
\end{align}
The function $\Treef(x; \Tree, \sM)$ is a \emph{regression tree} corresponding to the decision tree $\Tree$ and predictions $\sM$. A schematic of a single decision tree is given in Figure~\ref{fig:treefig}. BART was introduced by \citet{chipman2010bart} and, starting from the seminal work of \citet{hill2011bayesian}, has seen steadily increasing usage in causal inference. For recent reviews of BART see \citet{linero2017review} and  \citet{hill2019bayesian}.

The prior distribution for the decision trees $\Tree_t$ in BART encourages each tree to be \emph{shallow} in the sense that they are a-priori unlikely to extend beyond (say) depth $2$ or $3$. Because a single shallow decision tree will not be capable of modeling the true $g_0(x)$ on its own, each tree is thought of as a \emph{weak learner}. The strategy of adding many weak learners together to form a single flexible model is the same strategy employed by \emph{boosting algorithms} \citep{freund1999short,friedman2001greedy}.

The use cases of BART and Gaussian processes are essentially the same; indeed, we note that after conditioning on the tree structures, the BART model \emph{is} a Gaussian process. Specifically, given $(\Tree_1, \ldots, \Tree_T)$ we have $g \sim \GP(m, K)$ where $m(x) \equiv 0$ and $K(x, x') = \sigma^2_g \, N(x,x')$, where $N(x,x')$ is the proportion of trees where $x$ and $x'$ are associated to the same leaf node and $\sigma^2_g$ is the prior variance of the leaf node parameters. Hence, one possible view of BART is as a Gaussian process where the covariance function $K(x,x')$ is learned nonparametrically. As noted by \citet{linero2017review}, this relationship also holds in the limit as $T \to \infty$, with $g(x)$ converging (unconditionally )to a Gaussian process with a ``Laplace-like'' covariance.

The main difference from a modeling perspective is that BART priors induce shrinkage towards \emph{additive} models \citep{linero2017abayesian, rockova2017posterior}. For example, if we add together trees of depth no greater than 2 then the resulting $g(x)$ will have only second order interactions, whereas a Gaussian process with (say) the squared-exponential kernel will always have interactions of all orders. In practice it is often expected that the true $g_0(x)$ will be approximately additive, primarily consisting of main effects and perhaps some weaker second or third order interactions. It is this structure that BART models are designed to capitalize on, and this provides some explanation for BART's success in causal inference \citep{dorie2019automated}. A secondary difference is that BART models take $g(x)$ to be a step function, which can make BART inappropriate in settings where one wants to take advantage of smoothness in $g(x)$ (such as regression discontinuity designs); a modification of BART which is smooth is given by \citet{linero2017abayesian}. BART also has the computational advantage of not requiring elaborate approximations to be applied to large datasets. 

The two most common BART models are the \emph{semiparametric regression model}
\begin{math}
    Y_i \indep \Normal\{g(X_i), \sigma^2\}
\end{math}
and the \emph{nonparametric probit model}
\begin{math}
    Y_i \indep \Bernoulli[\Phi\{g(X_i)\}].
\end{math}
BART has also been extended to a large number of other models including heteroskedastic regression \citep{pratola2017heteroscedastic}, survival analysis \citep{sparapani2016nonparametric, basak2020semiparametric, linero2021bayesian, henderson2020individualized}, 
\ifarxiv conditional density estimation \citep{li2020adaptive}, \fi
log-linear models \citep{murray2020log}, and gamma regression \citep{linero2018shared}. Hence, for any given causal inference problem, there is likely an applicable BART model.

Implementing BART is a difficult task, and we refer the reader to \citet{chipman2010bart}, \citet{kapelner2014bartmachine}, and \citet{hill2019bayesian} for details on the computations. Fortunately, there are many \texttt{R} packages which fit BART and its variations. We note in particular the highly-optimized \texttt{bartMachine} and \texttt{dbarts} packages, the less-optimized but very broad \texttt{BART} package, and the \texttt{SoftBart} package which implements a variant of BART which produces continuous (rather the piecewise-constant) estimates \citep{linero2017abayesian}. We also note the \texttt{XBART} method of \citet{he2021stochastic}, which serves both as a fast approximation to BART for large datasets and as a method for quickly initializing BART MCMC samplers.

\subsection{The Bayesian Bootstrap}
\label{sec:bayesian-bootstrap}

Many causal estimands, such as average treatment effects (ATEs), are defined with respect to the population distribution of the confounding variables $X_i \iid F_X$. Frequentist methods usually implicitly approximate $F_X$ with its empirical distribution $\mathbb F_n = N^{-1} \sum \delta_{X_i}$, where $\delta_x$ is a point-mass distribution at $x$. By contrast, Bayesians are obligated to specify a prior for $F_X$, which is potentially a difficult problem when the dimension $P$ of $X_i$ is large.

The \emph{Bayesian bootstrap} \citep{rubin1981bayesian} bypasses this problem in a way that is similar to estimating $F_X$ with $\mathbb F_n$ while also accounting for uncertainty. The Bayesian bootstrap models $X_i$ with a discrete distribution which is supported at exactly the realized values of the covariates $(x_1, \ldots, x_J)$ so that $\Pr_\theta(X_i = x_j) = \omega_j$. We then specify an improper prior $\pi(\omega) = \prod_{j =  1}^J \omega_j^{-1}$. Conveniently, the posterior distribution is available as $[\omega \mid \bX] \sim \Dirichlet(m_1, \ldots, m_J)$ where $m_j$ is the number of times $x_j$ is realized in the sample. Using the Bayesian bootstrap is very simple in practice: for each sample of the outcome model $f_\theta(y \mid x)$ collected using MCMC, collect a matching sample of $\omega$ by sampling from the $\Dirichlet(m_1, \ldots, m_J)$ distribution.

The Bayesian bootstrap derives its name from its similarity to bootstrap resampling. When performing inference by MCMC, the Bayesian bootstrap bases inference on many samples of $F_X = \sum_{j = 1}^J \omega_j \, \delta_{x_j}$. By comparison, bootstrap resampling \citep{efron1993introduction} bases inference on samples of $F_X = \sum_{j=1}^J \omega_j \, \delta_{x_j}$ but with the Dirichlet posterior for $\omega$  replaced with $N\omega \sim \Multinomial(N, m_1 / N, \ldots, m_J / N)$.



When $X_i$ is a vector of confounders, we recommend using the Bayesian bootstrap as a prior for $F_X$ whenever possible; there is simply no payoff in constructing a complicated model for a high-dimensional distribution when it is not required. This recommendation also has theoretical backing \citep{ray2020semiparametric}. A weakness of the Bayesian bootstrap is that, because it does not perform any smoothing, it does not readily allow for imputation of missing values. This makes it unsuitable both for modeling missing potential outcomes and when some components of $X_i$ are missing.

\subsection{Infinite Mixture Models}
\label{sec:imm}

The Bayesian bootstrap is unfortunately not useful when some components of $X_{i}$ are missing. For example, suppose we have $X_i = (X_{i1}, X_{i2})$ with $X_{i1}$ missing; then the conditional distribution of $X_{i1}$ given $X_{i2}$ is $\Pr_\theta(X_{i1} = x_{j1} \mid X_{i2} = x_{2}) = \omega_j \, I(x_2 = x_{j2}) / \sum_{j'} \omega_j \, I(x_2 = x_{j'2})$, i.e., the distribution of $X_{i1}$ is determined by matching with the $x_j$'s which agree with the realized value of $X_{i2}$. An immediate problem arises if $X_{i2}$ is continuous, as there will be no observations which match $X_{i2}$ exactly.

To deal with this issue, we need a flexible prior for $F_X$ which is also smooth enough to not require exact-matching to impute missing values. A generic approach to constructing such a prior is to express $F_X$ as a \emph{mixture model}
\begin{align}
  \label{eq:mix}
  f_\theta(x)
  =
  \sum_{k=1}^K \omega_k \, q(x \mid \vartheta_k),
\end{align}
where $\omega = \{\omega_k :1 \le k \le K\}$ is a probability vector (so that $\sum_k \omega_k = 1, \omega_k \ge 0)$ and $\{q(x \mid \vartheta): \vartheta \in \mathscr T\}$ is a parametric family for the $X_i$'s (say, a multivariate normal distribution). This model has a natural interpretation as a \emph{latent class} model: associated to a given individual $i$ there is a latent class indicator $C_i$ such that (i) $C_i \iid \Categorical(\omega)$ and (ii) given $C_i = k$, we have $X_i \indep q(x \mid \vartheta_k)$.

When $K = \infty$ we refer to \eqref{eq:mix} as an \emph{infinite mixture model} (IMM). One convenience of IMMs relative to finite mixtures is that they allow us to avoid specifying the number of latent classes; instead, the number of clusters is chosen adaptively by the model to match the complexity of the data.   
The most popular IMM is the \emph{Dirichlet process mixture model} \citep{escobarwest1995, sethuraman1994}, which uses a \emph{stick breaking} prior for $\omega$
\begin{align}
    \label{eq:dpm-stick}
    \omega_k = \omega'_k \prod_{\ell < k} (1 - \omega'_\ell)
    \quad\text{where}\quad
    \omega'_\ell \iid \Beta(1, \alpha).
\end{align}
The parameter $\alpha$, called the \emph{concentration parameter}, dictates the number of latent classes which occur in the data (note that at most $N$ latent classes can occur in the sample even if $K = \infty$); it can be shown, for example, that if $M$ is the number of realized latent classes in the sample, then $(M - 1) \sim \Poisson\big(\alpha \sum_{i=0}^{N-1} (\alpha + i)^{-1}\big)$ approximately. The $\vartheta_k$'s are typically assumed to be drawn iid from a prior density $H$ which we refer to as the \emph{base} distribution.

IMMs are highly expressive, and they can approximate essentially any distribution under the right conditions. For example, if we take $f_\theta(x) = \sum_{k=1}^\infty \omega_k \, \Normal(x \mid \mu_k, \sigma^2_k \Identity)$ then $f_\theta(x)$ will be able to approximate any reasonable density to arbitrary precision \citep{shen2013adaptive}. IMMs also deal naturally with missing values; for example, if $X_i$ is continuous and bivariate then we can impute a missing $X_{i1}$ by sampling from the conditional distribution
\begin{align}
    \label{eq:mix-fc}
    f_\theta(x_1 \mid x_2)
    =
    \sum_{k=1}^K 
    \frac{
        \omega_k \, \Normal(x_2 \mid \mu_{k2}, \sigma_k^2)
    }{
        \sum_{k'=1}^K \omega_{k'} \, 
        \Normal(x_2 \mid \mu_{k2}, \sigma_k^2)
    }
    \Normal(x_1 \mid \mu_{k1}, \sigma_k^2).
\end{align}
Formula \eqref{eq:mix-fc} applies equally well when $X_{i1}$ and $X_{i2}$ are themselves vectors, with $\sigma_k^2$ replaced by $\sigma_k^2 \, \Identity$. Note that \eqref{eq:mix-fc} resembles the usual kernel density estimate (KDE) of $f_0(x_1 \mid x_2)$ using normal kernels, where $\sigma_k$ is the bandwidth parameter for both $x_1$ and $x_2$ \citep{muller1996bayesian}. More generally, if we replace the covariance matrix $\sigma_k^2 \, \Identity$ with a covariance matrix $\Sigma_k$ then the conditional distribution is a \emph{mixture of linear regressions} of the form $f_\theta(x_1 \mid x_2) = \sum_{k=1}^K \varpi_k(x_2) \, \Normal(x_1 \mid b_{k0} + b_{k1} x_2, \sigma_{1k}^2)$ for some $(b_{k0}, b_{k1}, \sigma_{k1})$ depending on $(\mu_k, \Sigma_k)$.


It is also straight-forward to include variables of mixed types in an IMM. For example, if $X_{i1}$ is Bernoulli and $X_{i2}$ is continuous, we can take
\begin{math}
    q(x \mid \vartheta_k)
    =
    v_k^{x_1} \, (1 - v_k)^{1 - x_1} \, \Normal(x_2 \mid \mu_k, \sigma_k).
\end{math}
Interpreted as a latent class model, this takes $X_{i1}$ and $X_{i2}$ to be independent within class $k$, with $X_{i2}$ normally distributed \citep{canale2015bayesian}.

Sampling from the posterior distribution of an IMM using the stick-breaking prior \eqref{eq:dpm-stick} is typically carried out using a \emph{Gibbs sampling} algorithm which truncates the model by setting  $\omega'_{K'} \equiv 1$ for some large $K'$. This strategy was initially explored by \citet{ishwaranjames2001}, who showed that $K'$ need not be taken very large to obtain a highly accurate approximation to $K = \infty$. We refer interested readers to \citet{ishwaranjames2001} and the other references in this section for implementation details. In the authors' experience, IMMs can often be implemented straight-forwardly using the \texttt{R} package \texttt{rjags} \citep{r-rjags} using the approximation $\omega \sim \Dirichlet(\alpha / K', \ldots, \alpha / K')$, which is similar to the stick-breaking approximation but easier to implement. There are several schemes for sampling from the posterior that bypass the need for truncation as well; these include the \emph{slice sampling} method of \citet{kalli2011slice} (which works for general IMMs) and the \emph{collapsed Gibbs sampling} approach discussed extensively by \citet{neal2000} (which works for Dirichlet process mixtures and, more generally, for IMMs that admit a \emph{\Polya\ urn representation}).


\subsection{Sparsity-Inducing Priors for High-dimensional Models}

We now consider the setting where the number of covariates $P$ is large relative to the sample size $N$, where some form of shrinkage or dimension reduction is required even if we are willing to assume a linear relationship between $X_i$ and $Y_i$. For simplicity, consider the classical linear model $Y_i = X_i^\top\beta + \epsilon_i$. In this scenario what we want is a Bayesian analog of the popular lasso \citep{tibshirani1996regression} and related methods.

A standard Bayesian approach to dealing with the large number of covariates is to assign independent \emph{spike-and-slab} prior distributions to each $\beta_j$ for $j=1, \dots, P$ \citep{mitchell1988bayesian, ishwaran2005spike, hahn2015decoupling}. Formally, this prior distribution is written as 
\begin{align*}
    [\beta_j \mid \gamma_j] \indep (1 - \gamma_j) \, \delta_0 + \gamma_j \, \Normal(0, \sigma_{\beta}^2)
    \qquad \text{and} \qquad
    \gamma_j \iid \Bernoulli(\tau),
\end{align*}
where $\delta_0$ is a point-mass distribution at $0$. The prior distribution for $\beta_j$ is a mixture between $\delta_0$ (the spike) and a normal distribution (the slab). The slab distribution need not be normal, and heavy-tailed distributions like the Cauchy have some theoretical advantages over the normal \citep{johnstone2005empirical}. The idea behind the spike-and-slab prior is that some covariates are important (therefore their corresponding coefficients are assigned a continuous prior distribution) while other covariates are superfluous (and can be removed from the model by having a zero coefficient). The indicator $\gamma_j$ tells us whether covariate $j$ is included in the model or not. In particular, one can look at posterior inclusion probabilities $\Pr(\gamma_j=1 \mid \bX, \bY)$ as a measure of variable importance. The prior inclusion probability $\tau$ dictates the degree of sparsity in the model and is typically assigned a beta prior distribution. This model works well when the true data generating process is sparse and only a small number of covariates have nonzero coefficients. Conversely, this model can also adapt to the dense setting by taking $\tau \approx 1$, which corresponds to a data-adaptive ridge regression estimator.

These spike-and-slab prior distributions are not unique to the linear regression setup. For example, we can easily assume an additive model of the form $Y_i = \sum_{j=1}^P g_j(X_{ij}) + \epsilon_i$. The problem is the same as before: we would like to remove unnecessary predictors $j$ for which $g_j(X_j) \equiv 0$. To this end, one can assign spike-and-slab priors $g_j \sim (1 - \gamma_j) \, \delta_0 + \gamma_j \, \GP(m_j, K_j)$ where $\GP(m, K)$ is a Gaussian process prior as described in Section~\ref{sec:GPs} and $\delta_0$ is now a point-mass at the zero function $0(x) \equiv 0$. This eliminates unnecessary predictors while keeping important predictors in the model in a way that does not assume any functional form for $g_j(\cdot)$ \citep{reich2009variable}. While this model removes linearity assumptions and induces sparsity, it still makes an additivity assumption on the effect of $X_i$ on $Y_i$. Less restrictive methods include the add-GP method of \citet{yang2015minimax} and the BART-based model of \citet{linero2016bayesian}, both of which allow for interactions between covariates as well as nonlinear relationships between $X_i$ and $Y_i$.

For settings where exact sparsity is not desired, continuous approximations to the spike-and-slab prior have also been proposed. Popular priors in this class include the horseshoe \citep{carvalho2010horseshoe} and Dirichlet-Laplace \citep{bhattacharya2015dirichlet} priors. The horseshoe prior, for example, sets
\begin{align*}
    [\beta_j \mid \lambda_j] \indep \Normal(0, \lambda_j^2) \quad \text{and} \quad
    \lambda_j \iid \Cauchy(0, v)
\end{align*}
where $\Cauchy(0,v)$ denotes a Cauchy distribution with scale $v$. These priors do not enforce strict sparsity, as none of the coefficients are set exactly to zero; however, a key feature of them is that they assign a large amount of mass to \emph{neighborhoods} of 0 while having heavy tails. This has the effect of aggressively shrinking small effects towards zero while avoiding shrinking the coefficients corresponding to important covariates. For similar reasons as the spike-and-slab prior distribution, this leaves them well-suited to handle high-dimensional parameter spaces, particularly when the true data generating mechanism is approximately sparse.

\section{Estimands and identifying assumptions}
\label{sec:estimands}

We now review some standard causal inference concepts within the potential outcomes framework \citep{rubin1974}. For simplicity, we consider a single binary exposure $A_i$ taking values in $\{0,1\}$. Let $Y_i(a)$ denote the \emph{potential outcome} of the response that would have been observed if unit $i$ had received the exposure level $a$ and let $A_i$ denote the observed value of the exposure. Let $X_i$ denote a vector of pretreatment covariates and confounders. In total, our observed data is $(\bY, \bA, \bX)$ where $\bY = (Y_1, \ldots, Y_N)$, $\bA = (A_1, \ldots, A_N)$, and $\bX = (X_1,\ldots,X_N)^\top$ stacks the $X_i$'s as rows in a matrix. Additional notation required for dealing with more complex settings will be introduced as needed. The fundamental challenge is to draw conclusions about the distribution of $\{Y_i(0), Y_i(1)\}$ given only samples of $(Y_i, A_i, X_i)$.


There are a number of causal quantities $\psi(\theta)$, i.e. estimands, that could be of interest in a particular study. We list a few common choices here, but these are by no means exhaustive. Arguably the most common estimands are the \emph{population average treatment effect} (PATE) defined by $\psi(\theta) = \E_\theta\{Y_i(1) - Y_i(0)\}$ and the \emph{sample average treatment effect} (SATE) defined by $\psi_N(\theta) = N^{-1} \sum_i \{Y_i(1) - Y_i(0)\}$. Variations of this targeting the treated population (ATT, or average treatment effect on treated) can be defined as $\E_\theta\{Y_i(1) - Y_i(0) \mid A_i = 1\}$, with analogous definitions for the control population with $A_i=0$. Frequently it is also of interest to understand the heterogeneity of a treatment effect as some observed characteristics vary; the conditional average treatment effect (CATE) is defined by $\psi_x(\theta) = \E_\theta\{Y_i(1) - Y_i(0) \mid X_i = x\}$, which describes how the treatment effect varies by characteristics of the population. A related estimand is the average causal effect within a subgroup $\sC$ given by $\E_\theta\{Y_i(1) - Y_i(0) \mid X_i \in \sC\}$; relevant subgroups $\sC$ might be specified a-priori or can be learned from the data itself \citep{sivaganesan2017subgroup}. For binary outcomes, it may be more meaningful to consider relative risks such as $\E_\theta\{Y_i(1)\} / \E_\theta\{Y_i(0)\}$ (or odds ratios) rather than differences in means; analogous versions of the PATE and CATE are easily defined in these settings.

Causal effects on other aspects of the potential outcome distribution, such as the quantiles, may also be of interest. If we let $F_{\theta 1}(y)$ and $F_{\theta 0}(y)$ be the cumulative distribution functions for the potential outcomes under treatment and control, respectively, we can define the treatment effect on the $\alpha^{\text{th}}$ quantile as $\psi_\alpha(\theta) = F_{\theta 1}^{-1}(\alpha) - F_{\theta 0}^{-1}(\alpha)$ for $\alpha \in (0,1)$. This quantile effect can be understood as the change in the population-level quantile obtained from changing all units from untreated to treated.

Because we do not observe both $Y_i(0)$ and $Y_i(1)$ for any observations in the sample, none of these estimands are identifiable from the observed data without additional assumptions. The estimands listed above are related, however, in that they are all identified under the same set of assumptions. Below, $\sX$ denotes the support of $F_X$.
\begin{description}[leftmargin = 0em, style = unboxed]
  \item[SUTVA] For any assignment $\ba = (a_1, \ldots, a_N)$ of units to treatments, the potential outcome $Y_i(\ba)$ depends only on $a_i$, so that we can write $Y_i(\ba) = Y_i(a_i)$. Additionally, we observe $Y_i \equiv Y_i(A_i)$, i.e., $Y_i = A_i \, Y_i(1) + (1 - A_i) \, Y_i(0)$. 
  \item[Positivity] The propensity score $\Pr_\theta(A_i = 1 \mid X_i = x) = e_\theta(x)$ satisfies $0 < e_\theta(x) < 1$ for all $x \in \sX$.
  \item[Strong Ignorability] The potential outcomes $\{Y_i(0), Y_i(1)\}$ are conditionally independent of $A_i$ given $X_i = x$ for all $x \in \sX$, i.e., $[(Y_i(0), Y_i(1) \perp A_i \mid X_i = x]$ for all $x \in \sX$.
\end{description}
The first assumption (\textbf{SUTVA}) is the stable unit treatment value which states that the outcomes of one unit are unaffected by the treatments received by other units and that there are not multiple versions of the treatment; that $Y_i = Y_i(A_i)$ is also known as a \emph{consistency} assumption. The second assumption (\textbf{Positivity}, also known as an overlap assumption) requires that each subject has some chance of being treated/untreated. The final (and arguably strongest) assumption (\textbf{Strong Ignorability}) states that we have measured all common causes of the treatment and outcome. This assumption is crucial, as it allows us to use information from the treated or control group to learn about the entire population of interest. Together, these assumptions allow us to write the estimands of interest in terms of observable quantities that can be estimated from the data. They identify the conditional density of the potential outcomes as $f_\theta\{Y_i(a) = y \mid X_i = x\} = f_\theta(Y_i = y \mid A_i = a, X_i = x)$, which is sufficient to identify all estimands which depend only on the marginal distributions of the potential outcomes.
For instance, the PATE can be written as $\E_\theta\{Y_i(1) - Y_i(0)\} = \int \{ \E_\theta(Y_i \mid X_i=x, A_i=1) - \E_\theta(Y_i \mid X_i=x, A_i=0) \} \, f_\theta(x) \ dx$ and the CATE can be written as $\E_\theta\{Y_i(1) - Y_i(0) \mid X_i = x\} = \E_\theta(Y_i \mid X=x, A=1) - \E_\theta(Y_i \mid X=x, A=0)$. 

We have focused on estimands that are identifiable under ignorability, but we discuss extensions to other estimands with different identification assumptions in Section \ref{sec:OtherEstimands}. 

\section{Nonparametric Bayesian Causal Inference}
\label{sec:bayesian-causal-inference}

In this section we will describe how nonparametric prior distributions should be constructed to avoid common pitfalls, how to compute causal effects from the output of an MCMC algorithm, and how to adapt these ideas to handle high-dimensional confounders.

\subsection{Regularization Induced Confounding and Prior Dogmatism}
\label{sec:prior-dogmatism}

The key feature of Bayesian nonparametric and high-dimensional methods which allows them to work is that they implement regularization through an informative prior on the parameter $\theta$. For example, Gaussian process priors regularize functions so that they are smooth, while spike-and-slab priors regularize many of the regression coefficients towards $0$. Special care must be taken when applying Bayesian nonparametric methods --- or machine learning methods in general --- to causal inference due to the phenomenon of \emph{regularization induced confounding} (RIC, \citealp{hahn2018regularization}). RIC occurs when our use of regularization to control the complexity of the nuisance parameters, such as the function $g(x)$ in \eqref{eq:semiparametric}, indirectly regularizes important causal parameters.

To demonstrate the potential risks of uncontrolled regularization, consider the semiparametric causal model with a binary treatment $A_i \sim \Bernoulli\{e_\theta(X_i)\}$ and outcome model
\begin{align}
    \label{eq:semipar-causal}
    Y_i(a) = g(a,X_i) + \epsilon_i,
    \quad \epsilon_i \sim \Normal(0, \sigma^2).
\end{align}
We examine the prior distribution on the \emph{selection bias} of the PATE, $\Delta(a) = \E_\theta\{Y_i(a) \mid A_i = a\} - \E_\theta\{Y_i(a)\}$. In words, $\Delta(a)$ is the bias in estimating $\E_\theta\{Y_i(a)\}$ using a naive estimate that does not control for confounders, $\sum_i I(A_i = a) \, Y_i / \sum_i I(A_i = a)$. The fact that we expect $\Delta(a) \ne 0$ is precisely the reason that inference about the PATE is non-trivial. It may be surprising, then, to find that straight-forward applications of nonparametric Bayes can result in $\Delta(a)$ being heavily shrunk towards $0$. 

\begin{figure}
    \centering
    \includegraphics[width=.9\textwidth]{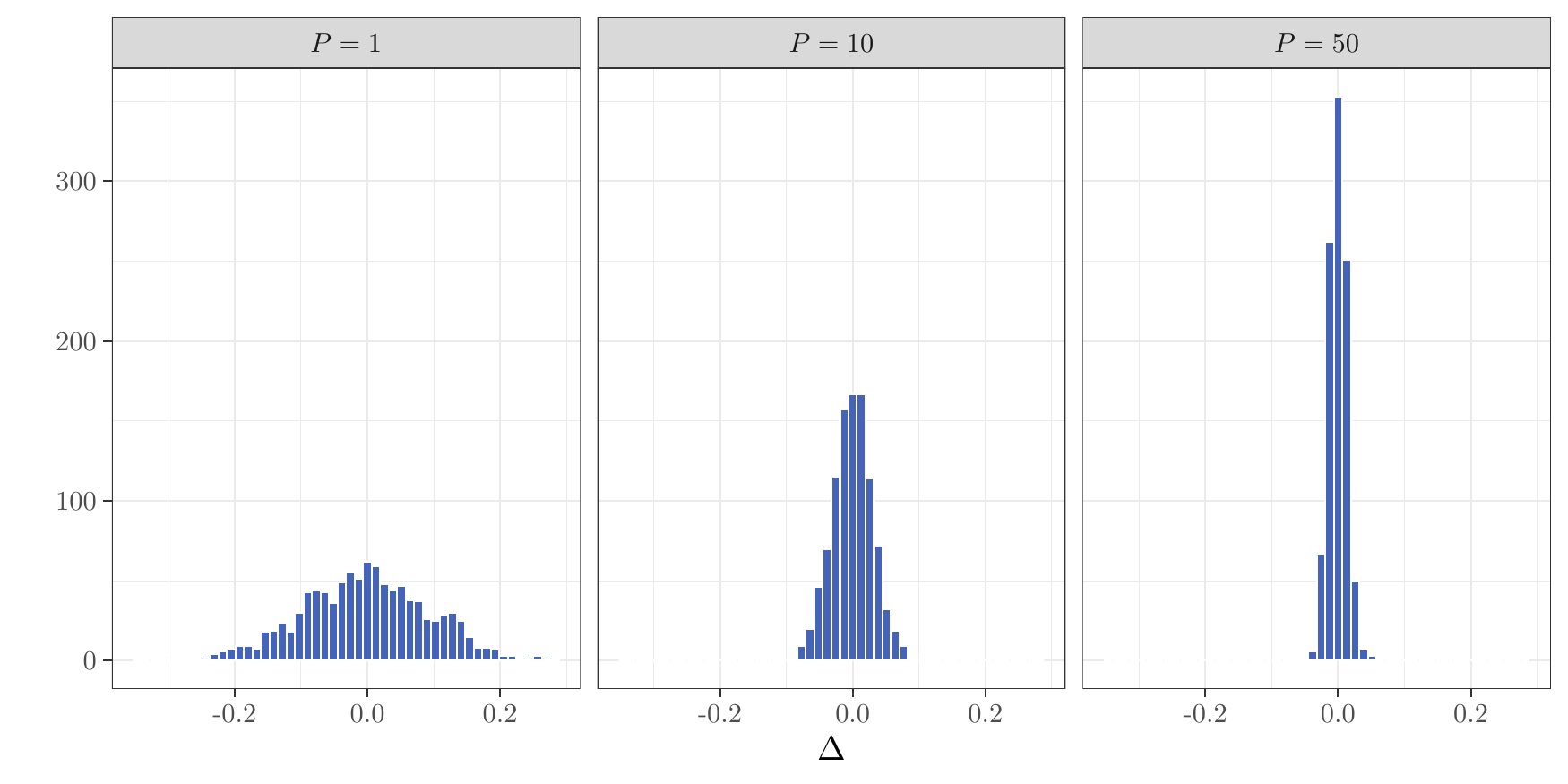}
    \caption{Prior distribution of the selection bias $\Delta(1)$ for the naive BART prior with $P \in \{1,10,50\}$}
    \label{fig:Concentration}
\end{figure}

\ifarxiv A detailed account of this phenomenon is given by \citet{linero2021high}. \fi
To gain intuition about why $\Delta(a)$ might be shrunk towards $0$, note that we can write
\begin{align}
    \label{eq:dogma}
    \Delta(1) = 
    \frac{\Var_\theta\{e_\theta(X_i)\}^{1/2} \, \Var_\theta\{g(1, X_i)\}^{1/2}}{\E_\theta\{e_\theta(X_i)\}} 
    \, \Cor_\theta\{e_\theta(X_i), g(1, X_i)\}
  .
\end{align}
When $e_\theta(x)$ and $g(1,x)$ are given independent priors it turns out that they are very likely to be nearly-uncorrelated if $P$ is of moderate dimension (or, in the case of parametric models, when $P$ is large). That independent random functions are ``probably approximately uncorrelated'' is an intrinsic property of many distributions on function spaces, with relatively narrow exceptions. Figure~\ref{fig:Concentration} shows the prior distribution of $\Delta(1)$ for a model which uses a BART prior for both the outcome regression $g(1,X_i)$ and the transformed propensity score $\Phi^{-1}\{e_\theta(x)\}$. As $P$ is increased we see that the prior distribution of $\Delta(1)$ concentrates sharply around $0$. Again, this behavior is not unique to BART, but is a general feature of nonparametric priors on high-dimensional spaces which do not model dependence between the selection and outcome models. 
\ifarxiv
\citet{linero2021high} shows that this also occurs for Gaussian process priors with the squared exponential kernel described in Section~\ref{sec:GPs}, with the particularly negative result that if the propensity score is bounded away from 0, i.e., $e_\theta(x) \ge \delta > 0$, then $\Delta(1) \sim \Normal(0, c^2)$ where $c \le \exp\{-CP\} / \delta^2$ for some constant $C$ independent of $e_\theta(x)$.
\else 
This also occurs for Gaussian process priors with the squared exponential kernel described in Section~\ref{sec:GPs}, with the particularly negative result that if the propensity score is bounded away from 0, i.e., $e_\theta(x) \ge \delta > 0$, then $\Delta(1) \sim \Normal(0, c^2)$ where $c \le \exp\{-CP\} / \delta^2$ for some constant $C$ independent of $e_\theta(x)$.
\fi

The inferential consequence of RIC is that the posterior distribution concentrates on models for which the selection bias is negligible, i.e., the data has no chance of overcoming the information about $\Delta(a)$ contained in the prior. Paradoxically, the act of including additional possible confounders makes the problem \emph{worse}. 
\ifarxiv
This is demonstrated in the experiments of \citet{hahn2020bayesian} and \citet{linero2021high} and theoretically studied by \citet{linero2021high} in the case of high-dimensional ridge regression.
\else
This is demonstrated in the experiments of \citet{hahn2020bayesian}.
\fi

Addressing RIC and prior dogmatism can be accomplished by introducing a functional relationship between the outcome model and the propensity score model. In order to force $e_\theta(X_i)$ and $g(1, X_i)$ to be correlated, we might set $g(1, x) = \mu(1,x) + \beta\{e_\theta(x)\}$ where the function $\beta(\cdot)$ is modeled nonparametrically. This approach was proposed by \citet{zhou2019penalized}, and it avoids prior dogmatism because $\beta(e)$ is a univariate function. Alternatively, \citet{hahn2020bayesian} incorporate the propensity score $e_\theta(x)$ as a predictor directly in $g(\cdot)$ by setting $Y_i(a) = g\{a, x, e_\theta(x)\} + \epsilon_i$. While the incorporation of $e_\theta(x)$ as a predictor might seem  redundant, it encourages the model to explain both the selection and outcome models with the same features of $x$, and hence increases the model's willingness to attribute differences in treatment groups to the confounders rather than the treatment.

In practice, this particular variant of prior dogmatism is easy to deal with: we simply include an estimate of the propensity score $e_\theta(X_i)$  --- or an estimate of it obtained independently of $\bY$ --- into the outcome model nonparametrically. We note that there are some advantages to using an estimated (rather than true) propensity score: (i) it is computationally much simpler and (ii) it removes the risk of model feedback, wherein misspecification of the propensity score taints an otherwise correctly specified outcome model \citep{zigler2013model, robins1997toward}. While one might feel uneasy about using the point estimate $\widehat e(X_i)$ in the outcome model rather than $e_\theta(X_i)$, we remark that this approach \emph{is} coherent from the Bayesian perspective 
\ifarxiv
\citep{hahn2020bayesian, linero2021high}.
\else
\citep{hahn2020bayesian}.
\fi

\subsection{A Useful Reparameterization of the CATE}
\label{sec:BCF}

If our goal is to estimate the conditional average treatment effect (CATE) $\tau(x) = g(1, X_i) - g(0, X_i)$ then care must be taken to ensure that it is appropriately regularized. If we place independent priors on $g(1, x)$ and $g(0, x)$, then $\tau(x)$ will typically be rather unstable, and we will tend to find heterogeneous effects where there are none. \citet{hahn2020bayesian} propose the following parameterization to deal with both the RIC phenomenon and instability in the CATE:
\begin{align}
  \label{eq:bcf-model}
  Y_i(a) = \mu(X_i, \widehat e_i) + a\, \tau(X_i) + \epsilon_i(a),
\end{align}
where $\widehat e_i$ is an estimate of $e_\theta(X_i)$ (note the inclusion of $\widehat e_i$ to eliminate RIC as detailed in Section~\ref{sec:prior-dogmatism}). Hence, rather than specifying (say) independent Gaussian processes for $g(1, x)$ and $g(0, x)$, we instead specify independent priors for $\mu(x, e)$ and $\tau(x)$. This allows differing amounts of regularization to be applied to the prognostic effect of the confounders $\mu(x, e)$ and the treatment effect $\tau(x)$; generally, one expects that heterogeneity in the treatment effect should be small relative to the effect of the covariates, and hence $\tau(x)$ might be shrunk heavily towards a constant so that the model is ``shrunk towards homogeneity.''

In principle, any nonparametric prior (Gaussian processes, BARTs, etc.) can be specified for $\mu(x,e)$ and $\tau(x)$. See \cite{GP_LI2020discussion} for a discussion on the choice of prior and its subsequent impact on uncertainty quantification. \citet{hahn2020bayesian} propose, in particular, to model $\mu(x,e)$ and $\tau(x)$ using BART, and refer to the resulting model as a \emph{Bayesian causal forest} (BCF). BCFs improve upon the seminal work of \citet{hill2011bayesian}, who initially proposed a BART model of the form $Y_i(a) = \mu(a, X_i) + \epsilon_i(a)$ with $\mu(a,x)$ given a standard BART prior. To encourage homogeneity, the BART prior for $\tau(x)$ expresses a preference for \emph{empty} trees consisting only of a root. This is sensible, as the treatment effect will be homogeneous whenever all of the decision trees comprising $\tau(x)$ are empty. BCFs are state-of-the-art for estimating SATEs, PATEs, and CATEs \citep{dorie2019automated}. Additionally, it is straight-forward to modify \eqref{eq:bcf-model} to allow for other response types, such as binary outcomes. In Section~\ref{sec:applications} we use a BCF to estimate the causal effect of smoking on medical expenditures as mediated by smoking's effect on an individual's health.

\subsection{Implementation Details}
\label{sec:implementation}

In this section we show how to numerically obtain estimates and measures of uncertainty from the posterior. We describe a number of estimands of increasing difficulty in terms of implementation; inferences for other estimands typically follow by similar logic. We assume that $B$ posterior samples the parameters $\theta_1, \ldots, \theta_B$ have already been collected via MCMC and that the analyst is now tasked with using those samples to perform inference.

Let $\psi(\theta)$ denote the causal estimand of interest. If $\psi(\theta)$ is a closed-form function of $\theta$ then inference is straight-forward, as we can simply compute $\psi(\theta)$ from the samples of $\theta$. One such example is the CATE, defined as $\psi_x(\theta) = \E_\theta\{Y_i(1) - Y_i(0) \mid X_i = x\}$. Under the causal assumptions in Section~\ref{sec:estimands}, this quantity can be written as $\psi_x(\theta) = \E_{\theta}(Y_i \mid A_i=1, X_i = x) - \E_{\theta}(Y_i \mid A_i=0, X_i = x)$, which is immediately available if we use the BCF model \eqref{eq:bcf-model}. Another example is the SATE, where under the assumption that $\epsilon_i(a) \equiv \epsilon_i$ in (say) model \eqref{eq:bcf-model} the SATE is given by $N^{-1} \sum_i \tau(X_i)$.

The problem is more difficult when the causal estimand is not a closed-form function of $\theta$. The simplest example where this occurs is the PATE, which cannot be simplified beyond the expression
\begin{align}
  \label{eq:g-comp-ate}
  \psi(\theta) = \int \{\E_\theta(Y_i \mid A_i = 1, X_i = x) - \E_\theta(Y_i \mid A_i = 0, X_i = 1)\} \, f_\theta(x) \ dx.
\end{align}
While a BCF provides a posterior distribution for $\E_\theta(Y_i \mid A_i = a, X_i = x)$, there is no guarantee that we can integrate this against $f_\theta(x)$ in closed form. As described in Section~\ref{sec:bayesian-bootstrap}, this problem can be solved using the Bayesian bootstrap \citep{rubin1981bayesian}, which allows us to write
\begin{align}
  \label{eq:g-comp-bb}
  \psi(\theta)
  =
  \sum_{j = 1}^J \omega_j \left\{ \E_\theta(Y_i \mid A_i = 1, X_i = x_j) - \E_\theta(Y_i \mid A_i = 0, X_i = x_j) \right\},
\end{align}
where $\omega \sim \Dirichlet(m_1, \ldots, m_J)$, $(x_1, \ldots, x_J)$ are the unique values of the $X_i$'s, and $m_j$ is the number of times we observe $X_i = x_j$. Using \eqref{eq:g-comp-bb} we can directly obtain samples of $\psi(\theta_b)$ to use for constructing point estimates and intervals.

Expression~\eqref{eq:g-comp-bb} is convenient when the Bayesian bootstrap is applicable, but this is not always the case; for example, if there is missingness in the covariates, then we cannot apply the Bayesian bootstrap. It will often be the case that, while computing \eqref{eq:g-comp-ate} directly is difficult, we \emph{can} approximate the integral using another round of Monte Carlo. Given $N \, K$ samples $X_{ik}^\star \sim f_\theta(x)$, we have the approximation
\begin{math}
  \psi(\theta) \approx (NK)^{-1} \sum_{i,k} \{\E_\theta(Y_i \mid A_i = 1, X_i = X_{ik}^\star) - \E_\theta(Y_i \mid A_i = 0, X_i = X_{ik}^\star)\}.
\end{math}
Here, $K$ denotes a number of ``pseudo-datasets'' which are used to approximate the integral, and can be set as large as desired to make the Monte Carlo error negligible. Even for $K = 1$ this is often quite accurate as an approximation to \eqref{eq:g-comp-ate}, and it is also straight-forward to explicitly compute the Monte Carlo standard error of $\psi(\theta)$ to ensure that it is small relative to our uncertainty in $\psi(\theta)$ \citep{linero2021simulation}; we remark that, even in the worst case where the Monte Carlo error is large, inferences will generally be conservative. The strategy of using Monte Carlo to compute causal effects was proposed by \citet{robins1986} and has been used many times \citep{imai2010general,linero2021simulation,scharfstein2014global}. 

\begin{algorithm}
  \caption{Monte Carlo PATE Estimation\label{alg:agc-1}}
  \textbf{Input:} $\theta, N, K$
  \begin{algorithmic}[1]
    \For{$k = 1,\ldots, K$}
      \For{$i = 1,\ldots,N$}
        \State $X_{ik}^\star \sim f_\theta(x)$
        \State $\Delta_{ik} \gets \E_\theta\{Y_i \mid A_i = 1, X_i = X_{ik}^\star\} - \E_\theta\{Y_i \mid A_i = 0, X_i = X_{ik}^\star\}$
      \EndFor
    \EndFor
    \State $\widehat \psi \gets (NK)^{-1} \sum_{i,k} \Delta_{ik} $
    \State 
    \begin{math}
      \widehat s^2 \gets 
      \sum_{i,k} (\Delta_{ik} - \widehat \psi)^2 / \{NK(NK - 1)\}.
    \end{math}
    \State \Return $\widehat \psi, \widehat s^2$
  \end{algorithmic}
\end{algorithm}

We summarize the Monte Carlo integration approach in Algorithm~\ref{alg:agc-1}, which returns both the approximation and a measure of its Monte Carlo variance. Given $B$ samples $\theta_1, \ldots, \theta_B$ from the posterior distribution, we compute $\widehat \psi_1, \ldots, \widehat \psi_B$ and $\widehat s^2_1, \ldots, \widehat s^2_B$ and are satisfied with the quality of the approximation as long as $B^{-1} \sum_b \widehat s^2_b$ is small relative to the sample variance of the $\widehat \psi_B$'s.

The Monte Carlo integration approach is also applicable to non-mean parameters, such as quantile causal effects. For example, if we are interested in the causal effect on the median $\psi(\theta) = \med_\theta\{Y_i(1)\} - \med_\theta\{Y_i(0)\}$ then we can approximate $\psi(\theta)$ with
\begin{align}
  \label{eq:g-comp-med}
  \psi(\theta)
  \approx 
  K^{-1} \sum_{k = 1}^K \med\{Y^\star_{1k}(1), \ldots, Y^\star_{Nk}(1)\}
  - \med\{Y^\star_{1k}(0), \ldots, Y^\star_{Nk}(0)\},
\end{align}
where we sample $X_{ik}^\star \sim f_\theta(x)$ and $Y^\star_{ik}(a) \sim f_\theta(y \mid A_i = a, X_i = X_{ik}^\star)$. It is again straight-forward to compute the Monte Carlo error of this approximation. A difference between \eqref{eq:g-comp-ate} and \eqref{eq:g-comp-med} is that \eqref{eq:g-comp-med} will tend to have higher Monte Carlo error due to the fact that \eqref{eq:g-comp-ate} performs ``Rao-Blackwellization'' by replacing $Y_{ik}^\star(a)$ with its expected value whereas \eqref{eq:g-comp-med} does not; this may necessitate a larger value of $K$, with $K = 10$ generally being more than sufficient; in our experience, this leads to negligible computational overhead relative to the computations required to run the MCMC in the first place. 
\ifarxiv
More advanced techniques for reducing the Monte Carlo error are discussed by \citet{rene2021causal}. 
\fi
One possible algorithm for the median is given in Algorithm~\ref{alg:agc-2}.

\begin{algorithm}
  \caption{Monte Carlo Median Effect Estimation\label{alg:agc-2}}
  \textbf{Input:} $\theta, N, K$
  \begin{algorithmic}
  \For{$k = 1,\ldots, K$}
    \For{$i = 1,\ldots, N$}
      \State $X_{ik}^\star \sim f_\theta(x)$
      \State $Y_{ik}^\star(0) \sim f_\theta(y \mid A_i = 0, X_i = X_{ik}^\star)$
      \State $Y_{ik}^\star(1) \sim f_\theta(y \mid A_i = 1, X_i = X_{ik}^\star)$
    \EndFor
    \State $\widehat \psi^{(k)} \gets \med\{Y_{1k}^\star(1), \ldots, Y_{Nk}^\star(1)\} - \med\{Y_{1k}^\star(0), \ldots, Y_{Nk}^\star(0)\}$
  \EndFor
  \State $\widehat \psi \gets K^{-1} \sum_k \widehat \psi^{(k)}$
  \State $\widehat s^2 \gets \sum_k (\widehat \psi^{(k)} - \widehat \psi)^2 / \{K(K-1)\}$
  \State \Return $\widehat \psi, \widehat s^2$
  \end{algorithmic}
\end{algorithm}

\subsection{Extensions to High-Dimensional Settings}

Here we detail how to tailor high-dimensional models to obtain improved estimates of average treatment effects in settings where $P > N$. We focus on the average treatment effect, but similar ideas could be applied analogously for other estimands. High-dimensional settings require some form of shrinkage or variable selection to reduce the dimension of the parameter space, and as discussed in Section~\ref{sec:prior-dogmatism} this must be tailored to the causal problem at hand to avoid regularization induced confounding. Typical high-dimensional approaches focus on prediction accuracy rather than causal estimation, and a naive application of these ideas can lead to substantial finite sample bias of treatment effects \citep{belloni2014inference}. As in Section~\ref{sec:prior-dogmatism}, a natural remedy to this issue is to incorporate the treatment assignment into shrinkage or variable selection, which can be done in a variety of ways.

One way to reduce finite sample bias that is induced by regularization is to include the propensity score into an outcome regression model, from which treatment effects can be estimated. This involves specifying a model for both the treatment and outcome as
\begin{align*}
    h_1^{-1} \big\{ E_{\theta}(Y_i \mid A_i = a, X_i = x) \big\} &= \beta_0 + \beta_a \, a + \beta_e \, h_2^{-1}\{e_{\theta}(x)\} + \sum_{j=1}^P \beta_j \, x_j \\
    h_2^{-1}\{e_{\theta}(x)\} &= \alpha + \sum_{j=1}^P \alpha_j \, x_j.
\end{align*}
This model has been used extensively, and was studied within the high-dimensional Bayesian paradigm by \citet{zigler2013uncertainty}. Estimation of the PATE from this outcome model can proceed as in Section~\ref{sec:implementation}. While in principle the covariate adjustment term $\sum_{j=1}^P \beta_j \, X_j$ could adjust for all sources of confounding bias alone, the propensity score helps to adjust for any remaining imbalances --- particularly those caused by shrinkage of the $\beta_j$ coefficients. To induce further dependence between the outcome and treatment models we can use a \emph{shared} spike-and-slab prior of the form
\begin{align*}
    \beta_j &\indep (1 - \gamma_j) \, \delta_0 + \gamma_j \, \Normal(0, \sigma_{\beta}^2), \\
     \alpha_j &\indep (1 - \gamma_j) \, \delta_0 + \gamma_j \, \Normal(0, \sigma_{\alpha}^2).
\end{align*}
The binary variables $\gamma_j$ are shared between the treatment and outcome models, which means that the same set of covariates are included in both models. This modification ensures that confounders whose effect on the outcome can be ``explained away'' by the treatment (exclusion of which can induce substantial bias in the PATE estimation) will not be excluded from the model \citep{hahn2018regularization}. Naive approaches to penalizing or shrinking the outcome model that were designed for prediction would likely exclude these confounders in finite samples and would therefore perform poorly for estimating the PATE.


Similar ideas for linking the treatment and outcome models for variable selection or model averaging have been proposed. \citet{wang2012bac} estimate treatment effects from the outcome model and perform model averaging to reduce the dimension of $X_i$ in a way that borrows information from a propensity score model. As in \cite{zigler2013uncertainty}, they utilize spike-and-slab prior distributions for both the treatment and outcome models, but do not force the set of variables included in the two models to be the same. Instead they link the two models using an informative prior distribution. Let $\gamma_j^a$ and $\gamma_j^y$ be binary variables indicating covariate $j$'s inclusion into the treatment and outcome models, respectively. They use a joint prior distribution on $(\gamma_j^a, \gamma_j^y)$ that leads to the following prior odds of inclusion into the outcome model:
\begin{align*}
    \frac{\Pr(\gamma_j^y = 1 \mid \gamma_j^a = 1)}{\Pr(\gamma_j^y = 0 \mid \gamma_j^a = 1)} = \varpi, \quad \frac{\Pr(\gamma_j^y = 1 \mid \gamma_j^a = 0)}{\Pr(\gamma_j^y = 0 \mid \gamma_j^a = 0)} = 1.
\end{align*}
The tuning parameter $\varpi > 1$ determines how much to prioritize covariates in the outcome model that are also included in the treatment model. This leads to increased probabilities of inclusion into the outcome model when a covariate is also included in the treatment model. Similar priors that prioritize covariates associated with both the treatment and outcome have been developed in various contexts (see \citealp{wang2015accounting, antonelli2017guided, wilson2018model, antonelli2019high, papadogeorgou2020causal}). While most of these approaches specifically focus on estimation using an outcome model, these ideas have also been extended to doubly robust estimators \citep{cefalu2017model, antonelli2020causal}; see Section~\ref{sec:DRestimators} for a discussion of double robustness. 

The high-dimensional approaches considered so far have mostly focused on variable selection and model averaging when linking the treatment and outcome models through informative prior distributions. A different approach that is based on a clever reparameterization of the problem was proposed by \cite{hahn2018regularization}. They focus on continuous treatments and outcomes, and reparameterize the treatment and outcome models as 
\begin{align*}
    \E_{\theta}(Y_i \mid A_i=a, X_i=x) &= \beta_a \, (a - x^\top \beta_c) + x^\top \beta_d \\
    \E_{\theta}(A_i \mid X_i=x) &= x^\top \beta_c.
\end{align*}
By parameterizing the problem in this way, we can assign independent shrinkage prior distributions for both $\beta_c$ and $\beta_d$. Regularization induced confounding is reduced since $\beta_c$ is in both the treatment and outcome models, and therefore shrinkage is determined by the association of each covariate with both the treatment and outcome; an alternate explanation is that by regressing $Y_i$ on the \emph{residual} $A_i - X_i^\top\beta_c$, we are estimating the effect of $A_i$ that cannot be explained by the confounders.

\subsection{Relationship with Doubly Robust Estimators}
\label{sec:DRestimators}

Doubly robust estimators are a commonly used tool in causal inference to reduce the impact of model misspecification and improve inferences \citep{scharfsteinetal1999, bang2005doubly}; see \citet{daniel2014double} for a comprehensive review. The core idea of doubly robust inference is that only one of either a propensity score or outcome regression model needs to be correctly specified to obtain consistent estimates of treatment effects. All other things being equal, this is clearly a desirable property. Additionally, if one correctly specifies both the propensity score and outcome regression models, then faster rates of convergence can be obtained for doubly robust estimators over plug-in estimators that utilize a single regression model. Roughly speaking, if both the propensity score and outcome regression estimators converge at $N^{1/4}$ rates or faster, then doubly robust estimators can converge at the parametric $N^{1/2}$ rate. This is particularly powerful in the high-dimensional or nonparametric settings discussed in this manuscript, where parametric rates of convergence are typically not possible. Examples of works that utilize this property can be found in \cite{farrell2015robust, chernozhukov2018double, antonelli2020causal}.

The methods discussed in this manuscript share certain commonalities with traditional doubly robust estimators. We have stressed the importance of including an estimate of the propensity score in the outcome model, and while the motivation is somewhat different --- being primarily motivated by regularization-induced confounding and prior dogmatism --- we are often led to similar operating characteristics. For example, \citet{zhou2019penalized} show that including a nonparametric adjustment for the propensity score as suggested in Section~\ref{sec:prior-dogmatism} can provide a form of double robustness. One might be interested in combining the desirable features of doubly robust estimators with the nonparametric Bayesian modeling techniques discussed in this article. One difficulty is that doubly robust estimators are inherently Frequentist and are not likelihood-based; therefore, it is not clear how to incorporate Bayesian methodology into the estimation of propensity scores and outcome regression models when using doubly robust estimators. This extension was explored in \cite{antonelli2020causal}, who showed that incorporating posterior distributions of propensity scores and outcome regression models can lead to improved inferential properties. We refer interested readers to \cite{antonelli2020causal}, but the main idea is to combine the nonparametric bootstrap with posterior distributions of unknown parameters to provide a frequentist inferential procedure that accounts for uncertainty in parameter estimation and allows for nonparametric Bayesian modeling. This leads to an inferential procedure that is asymptotically valid when both models are correctly specified and contract at sufficiently fast rates, and is conservative in finite samples or under model misspecification.

\section{Different Estimands and Identification Strategies}
\label{sec:OtherEstimands}

Here we explore estimands other than average treatment effects, and look into identification strategies that do not rely on strong ignorability assumptions. Given the vastness of the literature here, we necessarily must omit many important topics. We consider mediation analysis and regression discontinuity designs in detail, as these are two very popular topics in causal inference that have been shown to work well with Bayesian nonparametrics. Other applications of Bayesian nonparametrics include to  
instrumental variable models \citep{wiesenfarth2014bayesian, adhikari2020nonparametric}, to principal stratification \citep{schwartz2011bayesian}, to high-dimensional panel data settings \citep{antonelli2020causal}, and to replace synthetic control methods with flexible model-based methods \citep{ben2021augmented}. 


\subsection{Mediation}
\label{sec:mediation}

\begin{figure}
    \centering
    \includegraphics[width=.5\textwidth]{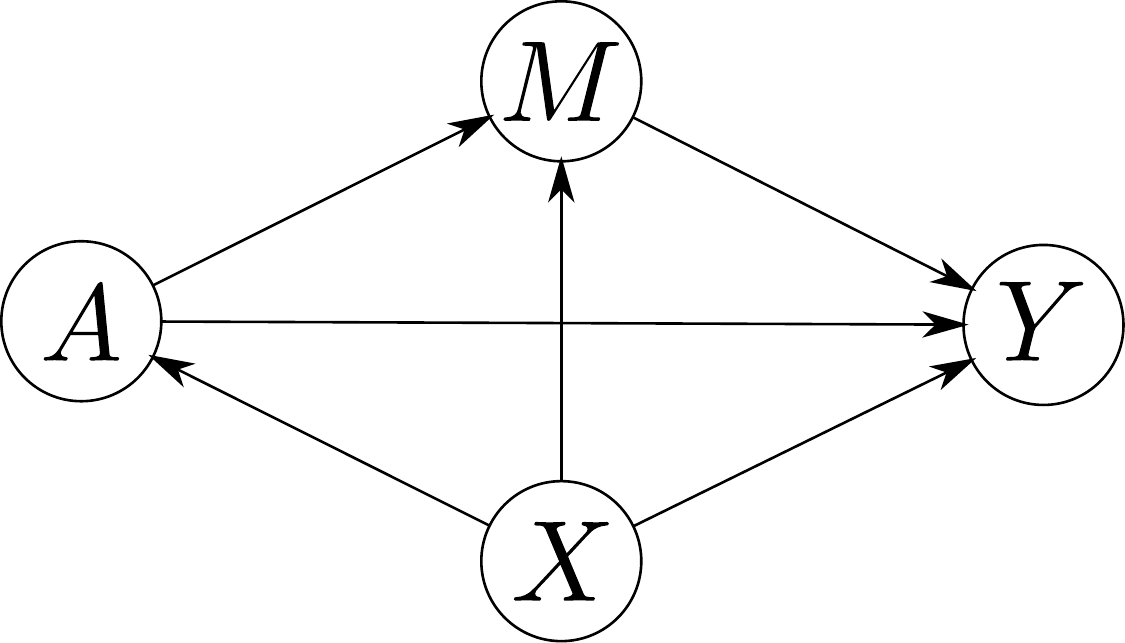}
    \caption{Directed acyclic graph depicting the causal structure underlying the simple mediation example of Section~\ref{sec:mediation}}
    \label{fig:mediation-fig}
\end{figure}

It is common in many settings for an exposure to influence the outcome along multiple distinct causal pathways. Causal mediation analysis considers the problem of decomposing the effect of an exposure into \emph{direct effects} (which correspond to the effect of an exposure directly on the outcome) and \emph{indirect effects} (which correspond to the effect of an exposure on the outcome which is due to its effect on an intermediate variable which also has a causal effect on the outcome). For a concrete example, \citet{imai2010general} consider the relationship between a job search skills seminar ($A_i = 1$ if someone attends the seminar and $A_i = 0$ otherwise) mediated by an individual's self-perceived self-efficacy in finding a job $(M_i(a))$ on that individual's depression at follow-up ($Y_i(a,m)$). So, for example, $M_i(1)$ is the job-search self efficacy for individual $i$ when they attend the seminar and $Y_i(0,m)$ is the depression of an individual who did not attend the seminar but is fixed to have job-search self-efficacy at the level $m$. We define the \emph{natural direct and indirect effects} for exposure level $a$ by
\begin{align*}
    \zeta(a)  = \E_\theta[Y_i\{1, M_i(a)\} - Y_i\{0,M_i(a)\}]
    \quad \text{and} \quad
    \delta(a) = \E_\theta[Y_i\{a, M_i(1)\} - Y_i\{a, M_i(0)\}]
\end{align*}
respectively. One approach to identifying $\zeta(a)$ and $\delta(a)$ is to make the \emph{sequential ignorability} assumption:
\begin{description}[style = unboxed, leftmargin = 0em]
  \item[SI1] The potential outcomes of the mediator and response are conditionally independent of the assigned treatment given measured confounders, i.e., $[\{Y_i(a',m), M_i(a)\} \perp A_i \mid X_i = x]$ for all $(a,a',x,m)$.
  \item[SI2] The potential outcomes of the response and mediator are conditionally independent given the assigned treatment and measured confounders, i.e., $[Y_i(a',m) \perp M_i(a) \mid A_i = a, X_i = x]$ for all $(a,a',x,m)$.
  \item[SI3] All combinations of the treatment and mediator are observable for all values of $x$, i.e., $f_\theta(A_i = a \mid X_i = x) > 0$ and $f_\theta\{M_i(a) = m \mid A_i = a, X_i = x\} > 0$ for all $(a,x,m)$.
\end{description}

The assumptions SI1 and SI2 are essentially variants of strong ignorability, and cannot be checked on the basis of the observed data, although in randomized studies SI1 may be known to hold a-priori. Under SI1 --- SI3 we can identify the marginal means $\mu(a,a') = \E_\theta[Y_i\{a, M_i(a')\}]$ as
\begin{align}
    \label{eq:g-med}
    \mu(a,a')
    =
    \int \E_\theta(Y_i \mid A_i = a, M_i = m, X_i = x) \, 
         f_\theta(M_i = m \mid A_i = a', X_i = x) \, f_\theta(x) \, dm \, dx.
\end{align}
Inference then boils down to (i) computing the posterior distribution of $\theta$ and (ii) computing \eqref{eq:g-med} and setting $\zeta(a) = \mu(1,a) - \mu(0,a)$ and $\delta(a) = \mu(a,1) - \mu(a,0)$. 

The step of computing \eqref{eq:g-med} is more complicated than computing the PATE in Section~\ref{sec:implementation} due to the fact that both the covariates and the mediator need to be integrated out. The Monte Carlo solution described in Section~\ref{sec:implementation} can still be applied, however, with one possibility being given in Algorithm~\ref{alg:agc-mediation}. Algorithm~\ref{alg:agc-mediation} has the advantage of being completely generic and working with missing confounders and covariates. A similar Monte Carlo integration strategy is proposed by \citet{imai2010general}.

\begin{algorithm}
  \caption{Monte Carlo Mediation Effect Estimation\label{alg:agc-mediation}}
  \textbf{Input:} $\theta, N, K$
  \begin{algorithmic}[1]
    \For{$k = 1,\ldots, K$}
      \For{$i = 1,\ldots,N$}
        \State $X_{ik}^\star \sim f_\theta(x)$
        \For{$a = 0, 1$}
        \State
          $M_{ik}^\star(a) \sim f_\theta(M_i \mid A_i = a, X_i = X_{ik}^\star)$
        \EndFor
        \For{$a \in \{0,1\} \text{ and } a' \in \{0,1\}$}
        \State $Y_{ik}^\star\{a, M_{ik}^\star(a')\} \gets
          \E_\theta\{Y_i \mid A_i = a, M_i = M_{ik}^\star(a'), X_i = X_{ik}^\star\}$
        \EndFor
      \EndFor
    \EndFor
    \For{$a = 0, 1$}
      \State $\widehat \delta(a) \gets (NK)^{-1} \sum_{i,k} [Y_{ik}^\star\{a, M_{ik}^\star(1)\} - Y_{ik}^\star\{a, M_{ik}^\star(0)\}]$
      \State $\widehat \zeta(a) \gets (NK)^{-1} \sum_{i,k} [Y_{ik}^\star\{1, M_{ik}^\star(a)\} - Y_{ik}^\star\{0, M_{ik}^\star(a)\}]$
    \EndFor
    \State \Return $\widehat \delta(0), \widehat \delta(1), \widehat \zeta(0), \widehat \zeta(1)$
  \end{algorithmic}
\end{algorithm}


\subsection{Regression Discontinuity Designs}

A classical situation in which treatment effects can be identified when the identifying assumptions of Section~\ref{sec:estimands} do not hold is the regression discontinuity design (RDD). See \citet{imbens2008regression} for a review of this design and corresponding statistical methods. An RDD occurs when there is a known relationship between treatment assignment and a pretreatment variable, $X_i$. We focus here on the sharp RDD, where the treatment assignment is defined as
$$
A_i = 
I(X_i \ge b).
$$
Here, $X_i$ is referred to as the running variable, and it completely determines the treatment assignment. A classic example of an RDD is one in which financial aid is given to students who meet a certain academic threshold \citep{van2002estimating}. One could also have a fuzzy RDD, which assume that $f_\theta(A_i=1 \mid X_i = x)$ has a discontinuity at $X_i=b$, though does not necessarily restrict these probabilities to be 1 or 0 as in the sharp RDD. Note here that the positivity assumption does not hold as treatment assignment is a deterministic function of the running variable. For this reason, estimating the PATE would rely heavily on model extrapolation. To avoid this, RDDs focus on local average treatment effects at the boundary $b$ given by $\psi(\theta) = \E_\theta\{Y_i(1) - Y_i(0) \mid X_i = b\}$.

One strategy to identify the local average treatment effect is to appeal to continuity of the CATE surface by assuming
\begin{align*}
  \E_{\theta}\{Y_i(a) \mid X_i=b\}
  = \lim_{x \rightarrow b} \E_{\theta}\{Y_i(a) \mid X_i=x\} 
\end{align*}
for $a \in \{0,1\}$. Under the sharp RDD, this identifies the treatment effect at the boundary as
\begin{align*}
  \psi(\theta)
  =
  \lim_{x \downarrow b} \E_\theta(Y_i \mid X_i = x) - \lim_{x \uparrow b} \E_\theta(Y_i \mid X_i = x).
\end{align*}
This identification strategy has led to estimating effects in RDDs using local regression approaches that assign more weight to individuals close to the boundary $b$. Most approaches have been shown to have poor inferential performance in terms of interval width and coverage, which has led to the development of more robust approaches that include bias-correction and confidence interval inflation to account for uncertainty in the estimated bias \citep{calonico2014robust}. While this has generally improved inference in RDDs, a recent approach using Bayesian nonparametrics has been shown to work quite well both in terms of estimation error and uncertainty quantification. 
\ifarxiv
  A natural way to estimate $\lim_{x \rightarrow b} \E_{\theta}(Y_i \mid X_i=x)$ using Bayesian nonparametrics is with a Gaussian process (see \citealp{chib2014nonparametric} for a spline-based Bayesian alternative). 
\else
  A natural way to estimate $\lim_{x \rightarrow b} \E_{\theta}(Y_i \mid X_i=x)$ using Bayesian nonparametrics is with a Gaussian process. 
\fi

When making predictions at the boundary a Gaussian process will naturally assign more weight to individuals close to the boundary, with weights dictated by the kernel function. Specifically, this approach models the potential outcomes as
\begin{align}
  \label{eq:rddgp}
  Y_i(a) = g_a(X_i) + \epsilon_i(a)
  \qquad\text{and}\qquad
  g_a \sim \GP(m_a, K_a).
\end{align}
The treatment effect in \eqref{eq:rddgp} is given by $\psi(\theta) = \mu_1(b) - \mu_0(b)$. One can construct point estimates and credible intervals directly from the posterior distribution of this quantity. Gaussian processes are particularly useful here for a number of reasons. For one, they make effectively no assumptions about the form of $\mu_a(x)$ and can consistently recover $\psi(\theta)$ under very mild conditions. They also naturally weigh the observations closer to $b$ more heavily, and the degree of this can be determined from the data by placing a hyperprior on $K_a$. Lastly, this process provides strong uncertainty quantification that can lead to improved finite-sample properties. \citet{branson2019nonparametric} found that this approach performed as well or better than the state-of-the-art approaches to estimation in RDDs. In particular, Gaussian process methods had interval coverage almost always at or near the desired level. 
\ifarxiv
  This is important, as recent work has found that most existing methods that are used for RDD designs tend to understate uncertainty and provide anti-conservative inference \citep{stommes2021reliability}. 
\else
  This is important, as most existing methods that are used for RDD designs tend to understate uncertainty and provide anti-conservative inference. 
\fi
This work has been further extended to spatial RDD settings where the goal is estimation of the treatment effect curve on a two-dimensional boundary \citep{rischard2020school}. Gaussian processes are also perfectly suited to this setting, as they can immediately handle a bivariate running variable $X_i = (X_{i1}, X_{i2})$, such as latitude and longitude. 

\subsection{Handling Missing Covariates with Infinite Mixtures}

In order to estimate the population PATE one needs to estimate the distribution of the covariates $X_i$ over which to average.  Using a Bayesian bootstrap prior for $f_\theta(x)$ is a simple and robust solution when all of the covariates are measured. When some continuous covariates are missing, however, this solution no longer works because the Bayesian bootstrap does not share information across similar values of the covariates.

As a first idea, one might use an IMM of the form $f_\theta(x) = \sum_{k=1}^\infty \omega_k \, q(x \mid \vartheta_k)$ in place of the Bayesian bootstrap. If we decompose $X_i = (X_{io}, X_{im})$ where $X_{io}$ is the observed part and $X_{im}$ is the missing part, then the conditional distribution of $X_{im}$ is given by
\begin{align}
    \label{eq:mix-impute}
    f_\theta(x_m \mid x_o)
    =
    \sum_k \frac{\omega_k \, q(x_o \mid \vartheta_k)}{\sum_{k'} \omega_{k'} \, q(x_o \mid \vartheta_{k'})} \, q(x_m \mid x_o, \vartheta_k).
\end{align}
If the conditional distribution $q(x_m \mid x_o, \vartheta_k)$ is easy to sample from then $f_\theta(X_{im} \mid X_{io}, Y_i, A_i)$ can be sampled using MCMC by (i) sampling a latent class $k$ with probability proportional to $\omega_k \, q(x_o \mid \vartheta_k)$, (ii) sampling from $q(x_m \mid x_o, \vartheta_k)$, and (iii) accepting the sampled $X_{im}$ according to the Metropolis-Hastings acceptance probability $1 \wedge \{f_\theta(Y_i, A_i \mid X_i^{\text{new}}) / f_\theta(Y_i, A_i \mid X_i^{\text{old}})\}$ \citep{robert2010introducing}. We can then model $Y_i$ and $A_i$ with BART models.


The above approach is somewhat inconvenient in that it specifies a separate class of models for the outcome, treatment, and confounders. We might instead be tempted to specify a single \emph{joint} model for $(Y_i, A_i, X_i)$ of the form $f_\theta(y,a,x) = \sum_k \omega_k \, q(y,a,x \mid \vartheta_k)$. \citet{wade2014improving} argue, however, that modeling a joint distribution $f_\theta(y,a,x)$ with an IMM weakens the model's expressiveness about the variables of interest $(Y_i, A_i)$ at the expense of capturing the dependence structure of $X_i$. This led \citet{roy2018bayesian} to adopt the \emph{enriched mixture modelling} framework of \citet{wade2014improving}, which takes
\begin{align}
    \label{eq:nested}
    f_\theta(y, a, x)
    =
    \sum_{g=1}^\infty \omega_g \sum_{s = 1}^\infty \omega_{s \mid g} \, q(y \mid a, x, \vartheta_g) \, q(a, x \mid \lambda_{s \mid g}).
\end{align}
The model \eqref{eq:nested} nests two mixture models: associated to each latent class $g$, there is collection of secondary latent classes $s$. The key is that the distribution of $Y_i$ depends only on $g$ rather than $s$, as capturing the structure of the univariate $Y_i$ requires fewer mixture components than capturing the structure of the multivariate $(A_i, X_i)$. When $\omega_g$ and $\omega_{s\mid g}$ are given independent stick-breaking priors according to \eqref{eq:dpm-stick} this model is referred to as an \emph{enriched Dirichlet process}.

In practice, $X_i$ will often contain a combination of both discrete and continuous variables. \citet{roy2018bayesian} keep the model as simple as possible by assuming independence within each mixture component, taking
\begin{align*}
    q(a, x \mid \lambda) = q_A(a \mid \lambda^{(A)}) \, \prod_{p=1}^P q_p(x_p \mid \lambda^{(p)}).
\end{align*}
For example, if $X_{ip}$ is binary we might take $q_p(x_p \mid \lambda^{(p)}) = \Bernoulli(x_p \mid \lambda^{(p)})$ while if $X_{ip}$ is continuous we might take $q_p(x_p \mid \lambda^{(p)}) = \Normal(\lambda^{(p)}_1, \lambda^{(p)}_2)$. This ensures that missing values of $X_{ip}$ can be imputed by first sampling a latent cluster pair $(s,g)$ with probability proportional to
\begin{math}
    \omega_g \, \omega_{s \mid g} \, q(y \mid a, x, \vartheta_g) \, q(a, x \mid \lambda_{s\mid g})
\end{math}
and then approximately sampling any missing $X_{ip}$'s from the distribution proportional to $q(y \mid a, x, \vartheta_g) \, q_p(x_p \mid \lambda_{s \mid g}^{(p)})$, which can be carried out by slice sampling \citep{slicesampling}. 

One possible drawback of IMMs that invoke within-class independence is that they compress the dependence structure of the observables into a single discrete variable. In many settings, the dependence may instead be explained by several \emph{latent factors}; when this occurs, it may take a very large number of latent classes to adequately capture the dependence. Semiparametric latent factor models, such as the mixed-data Gaussian copula model proposed by \citet{murray2013bayesian}, offer an attractive alternative to infinite mixture models in these settings.

\section{Application: The Medical Expenditure Panel Survey}
\label{sec:applications}

The Medical Expenditure Panel Survey (MEPS) is a collection of surveys intended to provide information about healthcare use and medical expenditures in the United States. The MEPS contains information about annual medical expenditures as well as demographic variables (age, sex, etc.), risk factors (e.g., cholesterol levels, smoking history) and disease indicators (e.g., whether an individual has had a heart attack, stroke, or cancer).

\begin{figure}
    \centering
    \includegraphics[width=0.49\textwidth]{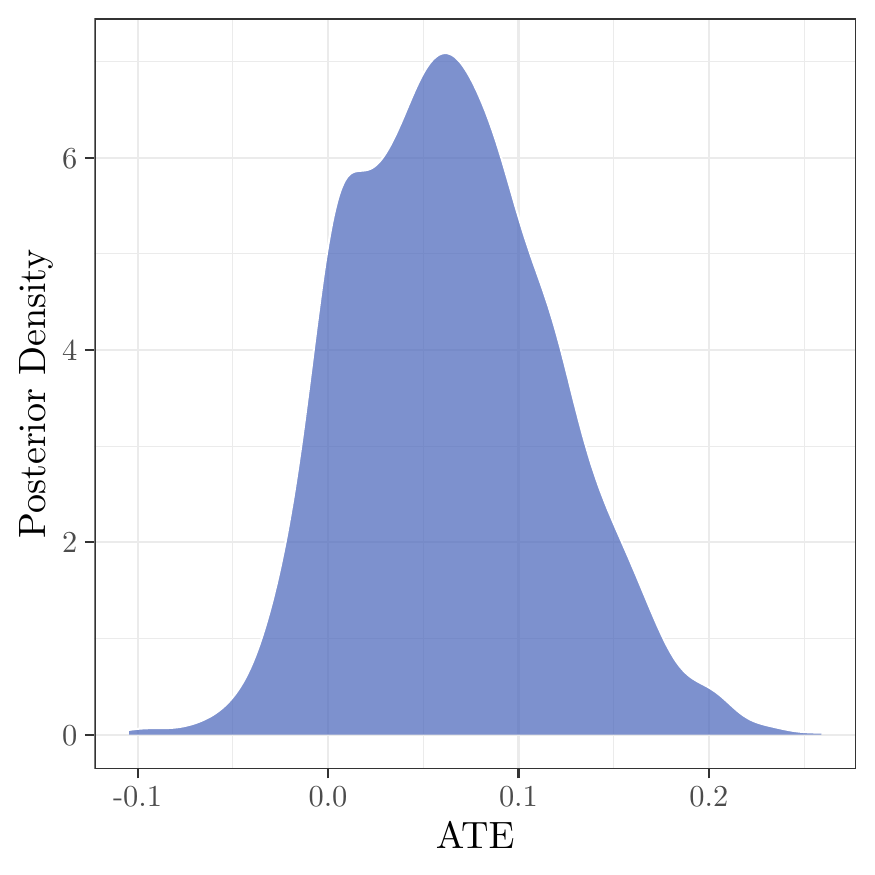}
    \includegraphics[width=0.49\textwidth]{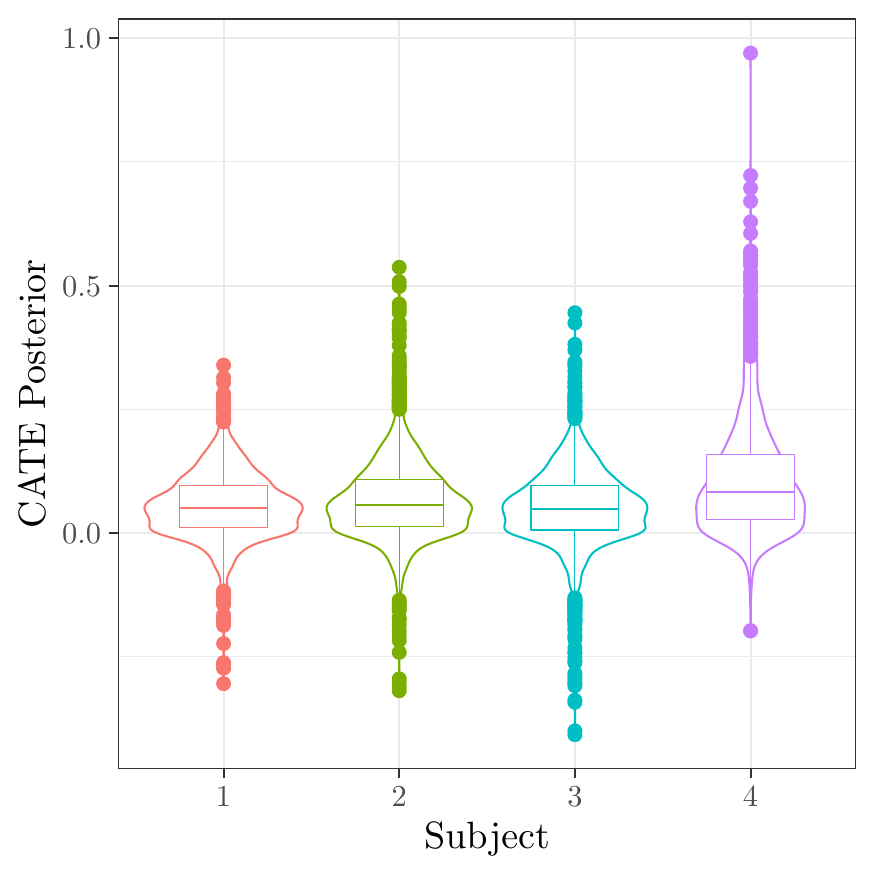}
    \caption{Left: posterior distribution of the average treatment effect of smoking for the MEPS dataset. Right: Posterior distribution of the conditional average treatment effect for four randomly chosen $X_i$'s.}
    \label{fig:MEPS}
\end{figure}

We illustrate the use of Bayesian causal forests for estimation of both the CATEs and PATE of smoking on total medical expenditures on the log scale after controlling for demographic variables (age, sex, income, education, poverty level) and risk factors (BMI, cognitive limitations) on a subset of roughly 10,000 women over the age of 18. For simplicity, we consider only individuals who incurred some medical expenditure, although in principle we could account for zero expenditures by modeling the probability of incurring some medical expenditure \citep{linero2018shared}. The posterior distribution of both the PATE and the CATE at four randomly chosen $X_i$'s are given in Figure~\ref{fig:MEPS}. These quantities were estimated using the Bayesian causal forest model described in Section~\ref{sec:BCF} and \eqref{eq:g-comp-bb} was used to compute the PATE using the Bayesian bootstrap. 

Summarizing the results of Figure~\ref{fig:MEPS}, there is evidence for a positive causal effect of smoking on medical expenditures. This is intuitive, as one would expect that smoking would lead to negative health outcomes, and hence to more use of the medical system. The estimated CATEs for each of four random individuals are largely in agreement with the PATE, although there seems to be some evidence of treatment effect heterogeneity, with the fourth subject being particularly different from the remaining three.

To study the hypothesis that smoking leads to higher medical expenditures due to its effect on health outcomes, we perform a mediation analysis using the self-perceived health status of an individual as a mediator; this serves as an omnibus measure of overall health, capturing many different health issues such as diabetes, cancer, and cardiovascular health. Individuals with lower perceived health have higher medical expenditures, so if smoking reduces overall health we might conjecture that smoking will have a positive indirect effect on expenditures mediated through perceived health. It is unclear what to expect from the direct effect, on the other hand --- it is possible, for example, that individuals who smoke may be less likely to seek medical care due to their tolerance for engaging in risky health behaviors.

 We let $M_i(a)$ denote an individuals' health for a given level of smoking ($a = 0$ for non-smoker and $a = 1$ for smoker) and $Y_i(a,m)$ denote the log medical expenditure for a given level of smoking $a$ and health $m$. We fit a the pair of models
\begin{align*}
    M_i(a) = r_m(a, X_i) + \nu_i(a), \\
    Y_i(a,m) = r_y(m, a, X_i) + \epsilon_i(a,m),
\end{align*}
where $r_m(a,x)$ and $r_y(m,a,x)$ are given independent BART priors. As with the BCF model, we control for regularization induced confounding by including the propensity for smoking as a predictor into both $r_m$ and $r_y$; additionally, we control for the estimates $\widehat r_m(0,X_i)$ and $\widehat r_m(1, X_i)$ in $r_y$ to control for regularization induced confounding on the path from the mediator to the outcome.

\begin{figure}
    \centering
    \includegraphics[width=.8\textwidth]{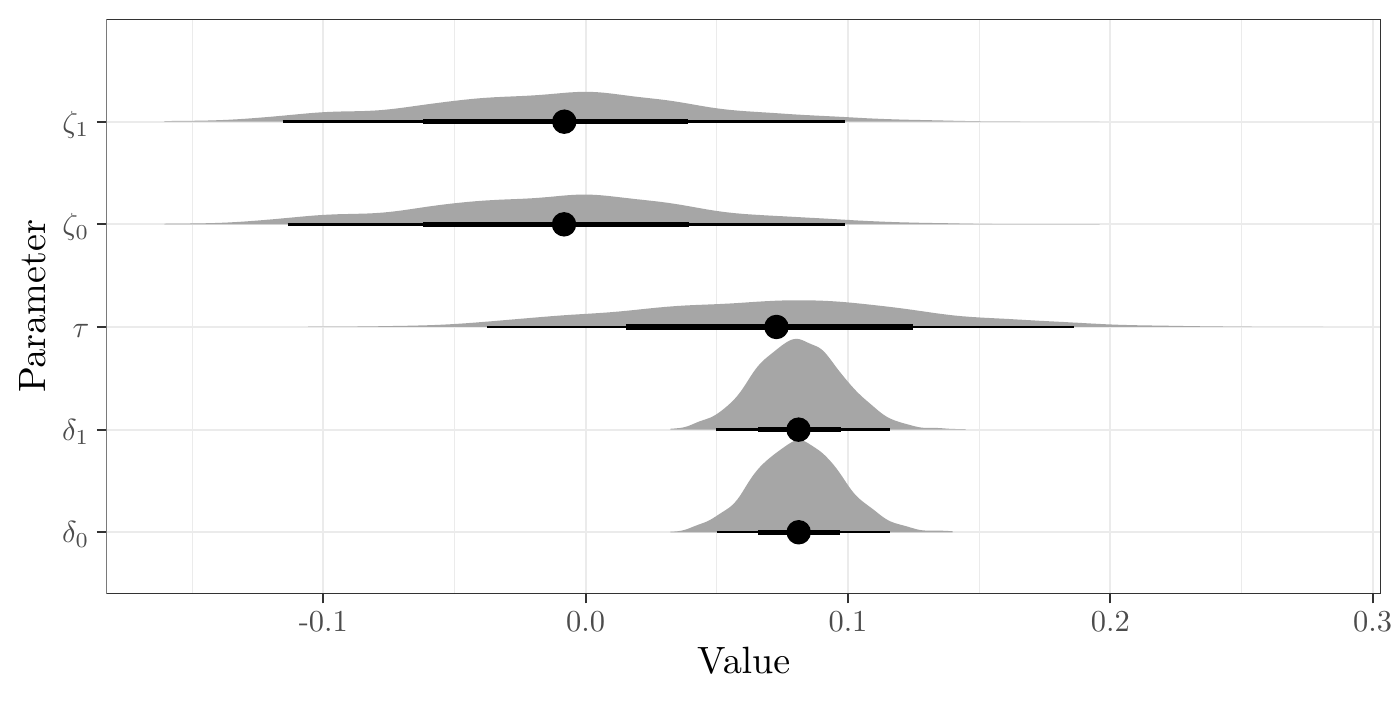}
    \caption{Direct, indirect, and total effects of smoking on medical expenditures.}
    \label{fig:MEPS-Mediation}
\end{figure}

Interval estimates for the direct and indirect effects are given in Figure~\ref{fig:MEPS-Mediation}. The results provide evidence for our conjecture; in particular, there is strong evidence of a positive indirect effect of smoking on medical expenditures. The evidence of a direct effect of smoking on medical expenditures is negligible, with the sign of the effect being unclear. The combination of these two effects results in a positive total effect of smoking on medical expenditures. In summary, by breaking the effect of smoking into direct and indirect effects, we are able to better tease out the causal effect of smoking on medical expenditures. To analyze the MEPS dataset we used the \texttt{bcf} package for PATE and CATE estimation and the \texttt{bartMachine} package to estimate the direct and indirect effects.

\section{Discussion}
\label{sec:discussion}

Bayesian nonparametric methodology provides elegant and powerful solutions for density and function estimation that naturally combine the modeling flexibility of modern machine learning with Bayesian uncertainty quantification. This combination of flexibility with uncertainty quantification is necessary to reduce model dependence and obtain valid estimates of causal quantities. In light of this, we believe that Bayesian nonparametrics and causal inference are a natural pairing, and the combination of the two paradigms is promising for both applied researchers and those developing new statistical methods.

Despite its potential benefits, Bayesian nonparametric methodology can be daunting to those who are unfamiliar with it, which can limit its use in applied research. With this article, we aim to bridge this gap by describing nonparametric Bayesian prior distributions in detail and carefully describing how they can be applied in causal inference contexts. While one must take care when using any nonparametric methodology for causal inference problems, we argue that for a wide class of problems estimation can easily be improved using standard nonparametric Bayesian prior distributions. 

Lastly, we emphasize that complex nonparametric Bayesian methodology is not an all-encompassing panacea for causal inference. The most important aspect of any causal analysis, particularly from observational data, is the plausibility of the assumptions needed to statistically identify the estimand, and Bayesian nonparametrics (or any statistical methodology) cannot help in this regard. Assessing sensitivity to identification is an important topic which we have avoided here. This topic will be treated in detail in the forthcoming monograph of \citet{daniels2021bayesian}.





\bibliographystyle{apalike}
\bibliography{mybib}

\end{document}